\documentclass[prd,aps,superscriptaddress,showpacs,preprint,preprintnumbers,amsmath,amssymb,nofootinbib]{revtex4}
\usepackage{graphicx}% Include figure files
\usepackage{dcolumn}% Align table columns on decimal point
\usepackage{bm}% bold math

\newcommand{\<}{\langle}
\renewcommand{\>}{\rangle}

\newcommand{\MSbar}{\overline{MS}}

\def\mres{m_{\rm res}}

\def\MeV{\rm MeV}
\def\LorInd{{\mu_1\mu_2\cdot\cdot\cdot\mu_n}}
\def\LorIndtwo{{\mu_2\cdot\cdot\cdot\mu_n}}
\def\Dcc{\stackrel{\,\leftrightarrow}{D}}

\def\mpi2{m_\pi^2}
\def\mK2{m_K^2}

\def\mres{m_{\rm res}}
\newcommand{\Dslash}{\rlap{/}\kern-2.0pt D}

\newcommand{\bea}{\begin{eqnarray}}
\newcommand{\eea}{\end{eqnarray}}
\newcommand{\be}{\begin{equation}}
\newcommand{\ee}{\end{equation}}

\newenvironment{outline}{
\noindent \framebox{Begin \ outline \hspace{5.0in} }
\begin{enumerate}}
{\end{enumerate}
\noindent \framebox{End \ outline \hspace{5.0in} }
\vspace{0.25in}}

\def\simge{%  ``greater than about'' symbol
    \mathrel{\rlap{\raise 0.511ex
        \hbox{$>$}}{\lower 0.511ex \hbox{$\sim$}}}}
\def\simle{%  ``less than about'' symbol
    \mathrel{\rlap{\raise 0.511ex
        \hbox{$<$}}{\lower 0.511ex \hbox{$\sim$}}}}

\begin{document}
\bibliographystyle{apsrev}

%%%%%%%%%%%%%%%%%%%%%%%%%%%%%% Counters %%%%%%%%%%%%%%%%%%%%%%%%%%
%%
%%  Counters to control sections printed and whether outline is
%%  printed.  Set value to 1 to print corresponding material
%%

\newcounter{Outline}
\setcounter{Outline}{1}

\newcounter{Intro}
\setcounter{Intro}{1}

\newcounter{StrucFunc}
\setcounter{StrucFunc}{1}

\newcounter{Renorm}
\setcounter{Renorm}{1}

\newcounter{LatticeSF}
\setcounter{LatticeSF}{1}

\newcounter{LatticeRenorm}
\setcounter{LatticeRenorm}{0}

\newcounter{Numerics}
\setcounter{Numerics}{1}

\newcounter{Results}
\setcounter{Results}{1}

\newcounter{Conclusions}
\setcounter{Conclusions}{1}

\newcounter{Acknowledgments}
\setcounter{Acknowledgments}{1}

\newcounter{Appendix}
\setcounter{Appendix}{0}

\newcounter{Tables}
\setcounter{Tables}{1}

\newcounter{Figures}
\setcounter{Figures}{1}

%%%%%%%%%%%%%%%%%%%%%%%%%%%%%% TITLEPAGE %%%%%%%%%%%%%%%%%%%%%%%%%%

\preprint{RBRC-475,CTP-3621}

\title{Nucleon structure functions with domain wall fermions}

\author{K.~Orginos}
\affiliation{CTP/LNS, Room 6-304,Massachusetts Institute of Technology, 
                 77 Massachusetts Av.,
                 Cambridge, MA 02139-4307}

\author{T.~Blum}
\affiliation{RIKEN-BNL Research Center, Brookhaven National Laboratory, Upton, NY 11973}
\affiliation{Physics Department, University of Connecticut, Storrs, CT 06269-3046}

\author{S. Ohta}
\affiliation{Institute for Particle and Nuclear Studies, KEK, Tsukuba, Ibaraki, 305-0801, Japan}
\affiliation{The Graduate University for Advanced Studies (SOKENDAI), Tsukuba, Ibaraki 305-0801, Japan}
\affiliation{RIKEN-BNL Research Center, Brookhaven National Laboratory, Upton, NY 11973}
%\author{A.~Soni}
%\affiliation{Physics Department, Brookhaven National Laboratory,Upton, NY 11973}

%\author{L.~Wu}
%\affiliation{Physics Department,Columbia University,New York, NY 10027}

\date{\today}

\begin{abstract}
We present a quenched lattice QCD calculation of the first few moments
of the polarized and un-polarized structure functions of the
nucleon. Our calculations are done using domain wall fermions and the
DBW2 gauge action with inverse lattice spacing $a^{-1}\approx$ 1.3
GeV, physical volume $V\approx (2.4$ fm)$^3$, and light quark masses
down to about 1/4 the strange quark mass ($m_\pi\approx 400$ MeV).
Values of the individual moments are found to be significantly larger
than experiment, as in past lattice calculations, but interestingly
the chiral symmetry of domain wall fermions allows for a precise
determination of the ratio of the flavor non-singlet momentum fraction
to the helicity distribution, $\<x\>_{u-d}/\<x\>_{\Delta u
- \Delta_d}$, which is in very good agreement with experiment. We
discuss the implications of this result. Next, we show that the chiral
symmetry of domain wall fermions is useful in eliminating mixing of
power divergent lower dimensional operators with twist-3
operators. Finally, we compute the isovector tensor charge at
renormalization scale $\mu=2$ GeV in the $\overline{MS}$ scheme,
$\<1\>_{\delta u - \delta d} = 1.192(30)$, where the error is the
statistical error only.

\end{abstract}

\maketitle

\newpage

%%%%%%%%%%%%%%%%%%%%%%%%%%%%%% INTRODUCTION %%%%%%%%%%%%%%%%%%%%%%%%%%

\section{Introduction}
\label{sec:intro}

\ifnum\theIntro=1
%%%%%%%%%%%%%%%%%%%%%%%%%%%%  Section %%%%%%%%%%%%%%%%%%%%%%%%%%%%%%%%
%
%  \section{Introduction}
%  \label{sec:intro}
%  File:  intro.tex
%
%%%%%%%%%%%%%%%%%%%%%%%%%%%%%%%%%%%%%%%%%%%%%%%%%%%%%%%%%%%%%%%%%%%%%%

\ifnum\theOutline=0
\begin{outline}
\item Talk about the importance of SF calculations
\item Review of what has been done.
\item Motivate the use of domain wall fermions
\item outline of the paper
\end{outline}
\fi

Quantum Chromodynamics (QCD) is the theory describing the strong
interactions, and hence it is responsible for the properties of
hadronic matter.  Unlike Quantum Electrodynamics (QED), its
non-perturbative nature, or strong coupling constant, 
makes it difficult to understand the low
energy content of the theory. The lattice formulation of QCD provides
both a non-perturbative way of defining the theory and a very powerful
tool to calculate its properties.

Deep inelastic scattering of leptons on nucleons has been the basic
experimental tool in probing
QCD~\cite{Breidenbach:1969kd,Friedman:1991nq,Kendall:1991np,Taylor:1991ew,Gluck:1995yr,Gehrmann:1995ag,Lai:1996mg,Adams:1997tq,Adeva:1997qz,Gluck:1998xa,Ackerstaff:1999ey,Martin:2001es}.
These experiments have given rise to the parton model and, through
extensive fits, have indirectly allowed the measurement of the parton
distribution functions, the basic structural blueprint for hadrons.
Connecting these experiments to the underlying theory of QCD is an
important theoretical endeavor.  During the last few years lattice
computations have provided many interesting results for nucleon matrix
elements~\cite{Gockeler:1996wg,Gockeler:1999jb,Dolgov:2002zm,Gockeler:2002mk,Orginos:2002mn,Hagler:2003jd,Ohta:2003ux,Ohta:2004mg,Sasaki:2003jh,:2003is,Gockeler:2003jf,Khan:2004vw},
in both quenched and full QCD. These calculations provide
first-principles values for the moments of structure functions at
leading twist.  One of the major unresolved
issues in these previous calculations is the approach to the chiral
limit; computational limitations have restricted calculations to
relatively large quark masses, introducing ambiguities in the
extrapolation to the chiral limit. Furthermore, values calculated
using the lattice regularization have significantly overestimated
results from fits to the experimental
data~\cite{Gockeler:1996wg,Dolgov:2002zm,Gockeler:2004wp}, leading to
suggestions in the literature that strong suppression in the chiral
limit is required to resolve the problem~\cite{Detmold:2001jb}.

We address this question with a calculation using domain wall
fermions~\cite{Kaplan:1992bt,Shamir:1993zy,Furman:1995ky}. Preliminary
results have been given
in~\cite{Orginos:2002mn,Ohta:2003ux,Ohta:2004mg}.  The use of domain
wall fermions allows us to examine the source of several systematic
errors.  Chiral symmetry at non-zero lattice spacing minimizes
discretization errors, $O(a^2)$ in this case.  Thus we work with
relatively coarse lattice spacing, $a\approx 0.15$ fm, and therefore
larger physical volume $L\approx 2.4$ fm. This means calculations with
light quark masses will not suffer unduly large finite size
corrections. In this study the lightest quark mass is roughly 1/4 of
the strange quark mass, as light as has been used in nucleon structure
calculations.  Chiral symmetry makes the renormalization properties of
operators simpler since there is less mixing with unwanted
operators. Here operators are non-perturbatively renormalized,
reducing a significant source of systematic error.  We have also
chosen to use the DBW2 gauge as it substantially reduces the already
small explicit chiral symmetry breaking for domain wall fermions with
finite extra fifth dimension~\cite{Orginos:2001xa,Aoki:2002vt}.

The remainder of this paper is organized as follows. In
Section~\ref{sec:StrucFunc} we briefly recall the polarized and
unpolarized structure functions of the nucleon arising from deep
inelastic scattering and the operators that arise from their operator
product expansions. The lattice transcription of operators and
correlation functions are described in
Section~\ref{sec:LatticeSF}. Perturbative and non-perturbative aspects
of operator renormalization are discussed in Section~\ref{sec:Renorm}.
Details of the numerical simulation are given in
Section~\ref{sec:Numerics}.  Section~\ref{sec:Results}, containing the
presentation and discussion of results, is the main part of the
paper. We summarize the present study and comment on future
calculations in Section~\ref{sec:conclusions}.

\fi

%%%%%%%%%%%%%%%%%%%%%%%%%%%%  Section %%%%%%%%%%%%%%%%%%%%%%%%%%%%%%%%

\section{Nucleon Structure Functions}
\label{sec:StrucFunc}

\ifnum\theStrucFunc=1
%%%%%%%%%%%%%%%%%%%%%%%%%%%%  Section %%%%%%%%%%%%%%%%%%%%%%%%%%%%%%%%
%
%  \section{Nucleon Structure Functions}
%  \label{sec:StrucFunc}
%  File:  StrucFunc.tex
%
%%%%%%%%%%%%%%%%%%%%%%%%%%%%%%%%%%%%%%%%%%%%%%%%%%%%%%%%%%%%%%%%%%%%%%

\ifnum\theOutline=0
\begin{outline}
\item How do we extract SF from the lattice
\end{outline}
\fi

The cross-section for deep inelastic scattering of leptons on a
nucleon target is given by the square of the matrix element for an
initial state lepton-proton pair to scatter to a final state of a
lepton and hadrons. After summing over all possible final states, the
square of the matrix element is computed using the optical theorem
which relates the summed, squared, matrix element to the forward
matrix element between nucleon states of the product of two
electromagnetic currents.
\begin{eqnarray}
\sigma & \sim & L^{\mu\nu}  W_{\mu\nu},\\
W_{\mu\nu} &=& i \int d^4 x e^{i q x} \langle N| T\{J^\mu(x), J^\nu(0)\}|N\rangle,\label{eq:W def}
\end{eqnarray}
where $L^{\mu\nu}$ and $W_{\mu\nu}$ are leptonic and hadronic tensors
respectively, and $q$ is the space-like four-momentum transfered to
the nucleon by scattering off the electron through a virtual photon.
The leptonic part is handled in perturbation theory since the QED
coupling constant is small; the hadronic part, however must be treated
non-perturbatively which is the focus of this paper.

The hadronic tensor is conveniently split, $W_{\mu\nu} = W^{[\mu\nu]} +  W^{\{\mu\nu\}} $.
The symmetric piece defines the unpolarized, or spin-average structure functions $F_1$ and $F_2$ (and $F_3$ if we consider neutrino scattering).
\begin{equation}
W^{\{\mu\nu\}}(x,Q^2) =  \left(-g^{\mu\nu} + \frac{q^\mu q^\nu}{q^2}\right)
 F_1(x,Q^2) \nonumber + \!
\left(p^\mu-\frac{\nu}{q^2}q^\mu\right)\!\!\!
\left(p^\nu-\frac{\nu}{q^2}q^\nu\right)\!
\frac{F_2(x,Q^2)}{\nu},
\end{equation}
while the anti-symmetric piece defines the polarized structure functions $g_1$
and $g_2$
\begin{equation}
W^{[\mu\nu]}(x,Q^2) = i\epsilon^{\mu\nu\rho\sigma} q_\rho
\left(\frac{s_\sigma}{\nu}(g_1(x,Q^2)+g_2(x,Q^2)) - 
\frac{q\cdot s p_\sigma}{\nu^2}{g_2(x,Q^2)} \right).
\end{equation}
$p_\mu$ and $s_\mu$ are the nucleon momentum and spin four-vectors,
$\nu = q\cdot p$, $s^2 = -m_N^2$ is our choice of normalization, $x=Q^2/2\nu$,  $Q^2=-q^2>0$ and $m_N$ is the nucleon mass.

At the leading twist in the operator product expansion of the two electromagnetic currents in Eq.~\ref{eq:W def}, the moments
of the structure functions can be factorized, at scale $\mu$, into
hard perturbative contributions (the Wilson coefficients)
and low energy matrix elements of local gauge invariant operators.
Adopting the notation of~\cite{Gockeler:1996mu},
\begin{eqnarray}
2 \int_0^1 dx x^{n-1} {F_1(x,Q^2)} 
&=& \sum_{q=u,d} c^{(q)}_{1,n}(\mu^2/Q^2,g(\mu))\: v_n^{(q)}(\mu),
\nonumber \\ 
\int_0^1 dx x^{n-2} {F_2(x,Q^2)} 
&=& \sum_{q=u,d} c^{(q)}_{2,n}(\mu^2/Q^2,g(\mu))\: v_n^{(q)}(\mu), 
\nonumber \\
2\int_0^1 dx x^n {g_1(x,Q^2)} 
  &=& \frac{1}{2}
\sum_{q=u,d} e^{(q)}_{1,n}(\mu^2/Q^2,g(\mu))\: a_n^{(q)}(\mu),\nonumber
 \\
2\int_0^1 dx x^n {g_2(x,Q^2)}
  &=& \frac{1}{2}\frac{n}{n+1} \sum_{q=u,d} [e^{(q)}_{2,n}(\mu^2/Q^2,g(\mu))
\: {d_n^{(q)}(\mu)} - \nonumber \\
  &-&  e^{(q)}_{1,n}(\mu^2/Q^2,g(\mu))\: a_n^{(q)}(\mu)]
\label{eq:Moments}
\end{eqnarray}
where $c^{(q)}_{i,n},e^{(q)}_{i,n}$ are the Wilson coefficients and
$ v_n^{(q)}(\mu), a_n^{(q)(\mu)}, {d_n^{(q)}(\mu)}$ are the non-perturbative
matrix elements. At the leading twist $ v_n^{(q)}(\mu)$ and $a_n^{(q)}$
are related to the parton model distribution functions 
$\langle x^n\rangle_q$ and $\langle x^n\rangle_{\Delta q}$:
\begin{eqnarray}  
  \langle x^{n-1}\rangle_q = v_n^{(q)} \;\;\;\;\;\;\;\;\;\;\;\;\;\;\;
  \langle x^{n}\rangle_{\Delta q} = \frac{1}{2} a_n^{(q)}
\end{eqnarray}

To determine $ v_n^{(q)}(\mu), a_n^{(q)}$, and ${d_n^{(q)}(\mu)}$
we need to compute non-perturbatively  the following matrix elements: 
\begin{eqnarray}
 \frac{1}{2} \sum_s \langle p,s|{{\cal O}^{q}_{\{\LorInd\}}}
 |p,s\rangle &=&
 2 v_n^{(q)}(\mu)\times
 [ p_{\mu_1}p_{\mu_2}
\cdot\cdot\cdot p_{\mu_n}+
\cdot\cdot\cdot -tr]\nonumber\\
 -\langle p,s|{{\cal O}^{5q}_{\{\sigma\LorInd\}}} |p,s\rangle &=&
 \frac{1}{n+1}a_n^{(q)}(\mu)\times
 [ s_\sigma p_{\mu_1}p_{\mu_2}\cdot\cdot\cdot p_{\mu_n}+\cdot\cdot\cdot
 -tr]\nonumber\\
 \langle p,s|{\cal O}^{[5]q}_{[\sigma\{\mu_1]\LorIndtwo\}}
 |p,s\rangle &=&
 \frac{1}{n+1}{d_n^{ q}}(\mu)\times
 [ (s_\sigma p_{\mu_1} - s_{\mu_1} p_{\sigma})p_{\mu_2}
\cdot\cdot\cdot p_{\mu_n}+\cdot\cdot\cdot -tr]\nonumber\\
\label{eq:matel}
\end{eqnarray}
$\{\}$ implies symmetrization and $[]$ anti-symmetrization of indices.
The nucleon states $|p,s\rangle$ are normalized so that
$\langle p,s|p',s'\rangle = (2\pi)^3 2E(p)\delta(p-p')\delta_{s,s'}$.
%and $s^2=-m^2_N$. 
The operators $\cal O$ are
\begin{eqnarray}
  {\cal O}^q_\LorInd  &=&
 \left(\frac{i}{2}\right)^{n-1}\bar{q}\gamma_{\mu_1} 
\Dcc_{\mu_2}\cdot\cdot\cdot \Dcc_{\mu_n}q -trace,\nonumber\\
{\cal O}^{5q}_{\sigma\LorInd} &=&
 \left(\frac{i}{2}\right)^{n}\bar{q}\gamma_{\sigma}\gamma_5 
 \Dcc_{\mu_2}\cdot\cdot\cdot \Dcc_{\mu_n}q -trace,
\label{eq:upol_pol_ops}
\end{eqnarray}
where $\Dcc = \overrightarrow{D} - \overleftarrow{D}$ and $\overrightarrow{D}$,
$\overleftarrow{D}$ are covariant derivatives acting on the right and the
left respectively.

In Drell-Yan processes the transversity distribution
$\langle x \rangle_{\delta q}$ can be measured 
(for details see~\cite{Jaffe:1991kp,Jaffe:1992ra,Barone:2001sp}).
The relevant matrix element is 
\begin{equation}
  \langle p,s|{{\cal O}^{\sigma q}_{\rho\nu\{\LorInd\}}}
 |p,s\rangle =
 \frac{2}{m_N}{\langle x^{n}\rangle_{\delta q}}(\mu)\times
 [(s_\rho p_{\nu} - s_{\nu} p_\rho)p_{\mu_1}p_{\mu_2}\cdot\cdot\cdot p_{\mu_n}
+\cdot\cdot\cdot -tr]\nonumber
\label{eq:matel_trans}
\end{equation}
where
\begin{equation}
{{\cal O}^{\sigma q}_{\rho\nu\LorInd}} =
\left(\frac{i}{2}\right)^{n}\bar{q}\gamma_5 \sigma_{\rho\nu} \Dcc_{\mu_1}\cdot\cdot\cdot \Dcc_{\mu_n}q - trace.
\label{eq:trans_ops}
\end{equation}

\fi

%%%%%%%%%%%%%%%%%%%%%%%%%%%%  Section %%%%%%%%%%%%%%%%%%%%%%%%%%%%%%%%

\section{Lattice Matrix Elements}
\label{sec:LatticeSF}

\ifnum\theLatticeSF=1
%%%%%%%%%%%%%%%%%%%%%%%%%%%%  Section %%%%%%%%%%%%%%%%%%%%%%%%%%%%%%%%
%
%  \section{Lattice Matrix Elements}
%  \label{sec:LatticeSF}
%  File:  LatticeSF.tex
%
%%%%%%%%%%%%%%%%%%%%%%%%%%%%%%%%%%%%%%%%%%%%%%%%%%%%%%%%%%%%%%%%%%%%%%

\ifnum\theOutline=0
\begin{outline}
\item Matrix element extraction
\item Simulation details 
\item Nucleon masses
\item Plateaus
\item fits
\end{outline}
\fi

The non-perturbative calculation of the matrix elements described in the previous section
(structure function moments) proceeds, as do all Euclidean lattice calculations, through the computation of
 nucleon three- and two- point correlation functions,
\begin{eqnarray}
  C^{\Gamma,\cal O}_{3pt}(\vec{p},t,\tau) &=& \sum_{\alpha,\beta}\Gamma^{\alpha,\beta}\langle J_{\beta}(\vec{p},t) {\cal  O}(\tau)\bar{J}_\alpha(\vec{p},0) \rangle,
\label{eq:3pt}\\
  C_{2pt}(\vec{p},t) &=& \sum_{\alpha,\beta}
\left(\frac{1+\gamma_4}{2}\right)_{\alpha\beta}
\langle J_{\beta}(\vec{p},t) \bar{J}_\alpha(\vec{p},0) \rangle,
\end{eqnarray}
where $\bar{J}(\vec{p},0)$ and $J(\vec{p},t)$
are interpolating fields
with the quantum numbers of the nucleon and definite momentum. $\Gamma^{\alpha\beta}$ is a Dirac matrix projection operator which is taken as  
\begin{equation}
\Gamma=\frac{1+\gamma_4}{2}
\end{equation}
for unpolarized matrix elements, and 
 \begin{equation}
  \Gamma=\frac{1+\gamma_4}{2}i\gamma_5\gamma_k,\;\;\; (k\ne 4)
 \end{equation}
 for polarized matrix elements. 
 $\frac{1+\gamma_4}{2}$ projects out 
 the positive parity part of the baryon propagator.
For the proton a typical choice for the interpolating field is 
\cite{Sasaki:2001nf}
\begin{equation}
  J_\alpha(\vec{p},t) = \sum_{\vec{x},a,b,c} e^{-i\vec{p}\cdot\vec{x}}\epsilon^{abc} \left[u^T_a(y_1,t) C \gamma_5 d_b(y_2,t)\right] u_{c,\alpha}(y_3,t) \phi(y_1-x)
 \phi(y_2-x)  \phi(y_3-x)
\end{equation}
with charge conjugation matrix $C=\gamma_4\gamma_2$, $\alpha$
a spinor index, and $a,b,c$ color indices. The functions $\phi(x)$
are smearing functions that are designed to maximize the overlap of
the interpolating field with the ground state of the nucleon. For the
source we used $\phi(x)=1$ when $x$ is within a box of size
$R\sim 1$ fm and zero outside this box. For the sink we took a point sink, $\phi(x)=\delta(x)$. We have optimized
the size of the box to maximize the overlap of the source to the ground
state. This setup works well for zero spatial momentum of the proton, 
and since we studied only this case so far, this was all we needed
to do. For non-zero momentum this smearing is not optimal; one must
resort to other smearing methods such as
gauge invariant Gaussian or Wupertal
smearing~\cite{Dolgov:2002zm,Gockeler:1996wg}. 

 In the limit when the Euclidean time separation between all operators is large, $t\gg \tau \gg 0$, the desired matrix element between ground states dominates the correlation function,
\begin{eqnarray}
 C_{2pt}(\vec{p},t) &=& Z_N \frac{E_N(\vec{p})+m_N}{2\,E(\vec{p})}
 e^{-E_N(\vec{p})t} + \cdots \nonumber\\
  C^{\Gamma,\cal O}_{3pt}(\vec{p},t,\tau) &=& Z_N \sum_{\alpha,\beta,s,s^\prime}
\Gamma_{\alpha\beta}\frac{u_\alpha(p,s) 
              \langle p,s | {\cal  O} |p,s^\prime \rangle \bar{u}_\beta(p,s^\prime) }{(2\,E(\vec{p}))^2}
	         e^{- E_N(\vec{p})t} + \cdots
\label{eq:asympt_3pt_2pt}
\end{eqnarray}
where $u(p,s)$ is the nucleon spinor satisfying the Dirac
equation, and $\langle 0| J_\alpha(\vec{p},t)|p,s\rangle = \sqrt{Z_N}
u_\alpha(p,s)$.  Using Eq.~\ref{eq:asympt_3pt_2pt} and
Eq.~\ref{eq:matel} (or Eq.~\ref{eq:matel_trans}) the desired matrix
elements can be extracted from the ratio of three point functions to
two point functions. In practice we would like to achieve the
asymptotic behavior of Eq.~\ref{eq:asympt_3pt_2pt} with as small as
possible $t$. For that reason the smeared interpolating
operator $J$ is essential. For more details on the technical aspects
of the lattice calculation the reader may refer
to~\cite{Martinelli:1989rr,Gockeler:1996wg,Gockeler:2000ja,Dolgov:2002zm}.

The momentum fraction $\<x\>_q$ carried by each valence quark in the nucleon is computed in Euclidean space by inserting into the correlation function the operator
 \begin{equation}
 {\cal O}^q_{44}  =  \bar{q}\left[\gamma_{4}\Dcc_{4}-
\frac{1}{3}\sum_k \gamma_{k}\Dcc_{k}\right]q.
 \end{equation} 
 $\Dcc$ is the lattice covariant derivative, or difference operator,
 \begin{equation}
 \Dcc \,= \overrightarrow{D} - \overleftarrow{D}
 \end{equation}
with 
\begin{eqnarray}
\overrightarrow{D}q = \frac{1}{2}\left[U_\mu(x)q(x+\hat\mu) -  U^\dagger_\mu(x-\hat\mu)q(x-\hat\mu)\right]\nonumber\\
\bar{q}\overleftarrow{D} = \frac{1}{2}\left[\bar{q}(x+\hat\mu)U_\mu^\dagger(x) -  \bar{q}(x-\hat\mu)U_\mu(x-\hat\mu)\right].
\end{eqnarray}
${\cal O}^q_{44}$ belongs to the ${\bf 3^+_1}$
 representation~\footnote{The representations of H(4) are denoted as
 ${\bf d^C_n}$, where ${\bf d}$ is the dimension of the
 representation, ${\bf C}$ is the charge conjugation and the subscript
 distinguishes between different representations of the same
 dimensionality and charge conjugation.}  of the hypercubic group H(4)
 and does not mix with any lower dimensional operators under
 renormalization~\cite{Mandula:1983us,Gockeler:1996mu}. The
 ratio, \begin{equation} R_{\langle x \rangle_q}
 = \frac{C_{3pt}^{\Gamma,{\cal O}^q_{44}}}{C_{2pt}} =
 m_N \<x\>_q, \end{equation} then yields the momentum fraction.

Similarly, the helicity distribution $\<x\>_{\Delta q}$ for each valence quark is computed from the operator
 \begin{equation}
 {\cal O}^{5q}_{\{34\}}  =  i \bar{q}\gamma_5
\left[\gamma_{3}\Dcc_{4}+\gamma_{4}\Dcc_{3}\right]q,
 \end{equation}
 belonging to the ${\bf 6^-_3}$
 representation of H(4). It also does not mix with
 lower dimensional operators under renormalization~\cite{Mandula:1983us,Gockeler:1996mu}. The ratio yields
 \begin{equation}
 R_{\<x\>_{\Delta q}} = \frac{C_{3pt}^{\Gamma,{\cal O}^{5q}_{\{34\}}}}{C_{2pt}} = m_N \<x\>_{\Delta q}.
 \end{equation}

The lowest moment of the transversity, $\<1\>_{\Delta q}$,  related to the tensor charge of the nucleon, 
is computed  in Euclidean space using the operator
 \begin{equation}
 {\cal O}^{\sigma q}_{34}  = \bar{q}\gamma_5 \sigma_{34}q,
 \end{equation}
with $\sigma_{\mu\nu} = \frac{i}{2}[\gamma_\mu,\gamma_\nu]$.
This operator belongs to the ${\bf 6^+_1}$
representation of H(4) and does not mix with any
lower dimensional operators under renormalization~\cite{Mandula:1983us,Gockeler:1996mu}. Again
 \begin{equation}
 R_{\<1\>_{\delta q}} = \frac{C_{3pt}^{\Gamma,{\cal O}^{\sigma q}_{34}}}{C_{2pt}} =\<1\>_{\delta q}.
 \end{equation}

Finally the twist-3 matrix element $d_1$ related to $g_1$ and $g_2$ is given by the operator 
\begin{equation} {\cal O}^{5 q}_{[34]}
 = i\bar{q}\gamma_5 
\left[\gamma_3\Dcc_4  - \gamma_4\Dcc_3\right] q,
\end{equation} 
belonging to
 the ${\bf 6^+_1}$ representation of H(4).
 ${\cal O}^{5 q}_{[34]}$ is allowed to mix with lower dimensional operator ${\cal O}^{\sigma q}_{34}$ if the lattice fermions do not respect chiral
 symmetry. The mixing coefficient in this case is linearly divergent
 with the inverse lattice spacing and hence a non-perturbative subtraction is required~\cite{Gockeler:1999jb}. The use of domain wall fermions eliminates this problem as first shown in \cite{Ohta:2003ux} and in Section~\ref{sec:Results}.  $d_1$ is given by the ratio
 \begin{equation} 
R_{d_1} = \frac{C_{3pt}^{\Gamma,{\cal O}^{5 q}_{[34]}}}{C_{2pt}} =d_1.
\end{equation}

\fi

%%%%%%%%%%%%%%%%%%%%%%%%%%%%  Section %%%%%%%%%%%%%%%%%%%%%%%%%%%%%%%%
\section{Renormalization}
\label{sec:Renorm}

\ifnum\theRenorm=1
%%%%%%%%%%%%%%%%%%%%%%%%%%%%  Section %%%%%%%%%%%%%%%%%%%%%%%%%%%%%%%%
%
%  \section{Renormalization}
%  \label{sec:Renorm}
%  File:  Renorm.tex
%
%%%%%%%%%%%%%%%%%%%%%%%%%%%%%%%%%%%%%%%%%%%%%%%%%%%%%%%%%%%%%%%%%%%%%%

\ifnum\theOutline=0
\begin{outline}
\item NPR renormalization conditions
\item Anomalous dimensions
\item MSbar to NPR matching
\end{outline}
\fi

The local operators discussed in the previous section arise from an operator product expansion and therefore must be renormalized. The renormalized operators defined at scale $\mu$ are obtained from 
lattice-regularized operators defined with lattice spacing $a$.
\begin{equation}
  {\cal O}_i(\mu) =  Z_i(\mu,a){\cal O}_i(a) +
\sum_{j\ne i} a^{d_j - d_i}Z_{ij}(\mu,a) {\cal O}_j(a),
\label{eq:renorm_gen}
\end{equation}
where ${\cal O}_j$ are a set of operators allowed by symmetries to
mix among themselves, and $d_j$ is the dimension of each operator. If mixing with lower
dimensional operators occurs, the mixing coefficients are power
divergent, and hence must be computed non-perturbatively to
accurately subtract them. The
mixing of lattice operators is more complicated than that of the
continuum operators since not all of the
continuum symmetries are respected on the lattice. In particular, $O(4)$ rotational symmetry in
Euclidean space is broken down to the hypercubic group $H(4)$.  As a
result, an irreducible representation of $O(4)$ is reducible under
$H(4)$ and hence mixing of operators that would not occur in the
continuum can occur on the lattice. For a detailed analysis of the
$H(4)$ group representations see~\cite{Mandula:1983us,Gockeler:1996mu}
and references therein.  The lattice operators are selected carefully so  mixing with lower dimensional
operators does not occur and thus no power divergences are encountered.  
In general, the breaking of rotational symmetry makes the calculation of higher spin operators (and thus higher moments of structure functions) difficult.

The breaking of chiral symmetry ({\it e.g.}, by the lattice) results in mixings with
lower dimensional operators for the $d_n$ matrix elements. The problem
is avoided by using chiral lattice fermions such as domain wall,
overlap, or fixed point fermions.  

In order to renormalize the quark bilinear operators studied here, we employ the non-perturbative
renormalization (NPR) method introduced in~\cite{Martinelli:1995ty}.  This method has been shown to work
very well in the case of domain wall fermions~\cite{Blum:2001sr,Dawson:2002nr}.
For quark bilinears without derivatives we only need to compute
a single quark propagator from a point source; the Fourier
transform then yields 
 the quark propagator ${\bf S}(pa;0)$ in momentum space which is sufficient to calculated all of the needed vertices.
\begin{equation}
{\bf S}(pa;0) = \sum_{x} e^{-ip \cdot x} {\bf S}(x;0),
\end{equation}
where the lattice momentum $p_\mu$ is
\begin{equation}
p_\mu = \frac{2\pi}{L_\mu} n_\mu~~~~~~(n_\mu = 0,\pm 1,\pm 2,\dots),
\end{equation}
where $L_\mu$ is the linear lattice size in direction $\mu$.
Because the NPR method relies on matrix elements of the operators between off-shell quark and gluon states, the calculation proceeds in 
a fixed gauge which, for convenience, 
is chosen to be the Landau gauge.

In the case of operators containing derivatives, a single quark
propagator is not sufficient to build the quark bilinears
we need. One way to proceed is the method
in~\cite{Gockeler:1998ye} where propagators with
momentum sources instead of point sources are prepared in Landau gauge and derivative (difference) operators are constructed at the sink
point where values at all neighboring points are available.
This method works well when operators with many derivatives are needed but is expensive if many values of momentum are desired. 

In our case we are only interested in operators with one derivative
which can be constructed from  a  point-source propagator and a point-split-source propagator with appropriate gauge links attached at the source.
\begin{equation}
{\bf S}_\mu(pa;0) = \sum_{x} e^{-ip \cdot x} \frac{1}{2}
\left[
   {\bf S}(x;-\hat\mu) U_\mu(-\hat\mu) -
   {\bf S}(x;\hat\mu) U^{\dagger}_\mu(0)
\right].
\end{equation}
Following this strategy,
we compute the matrix elements of derivative operators with many values of momentum by computing four additional quark propagators on each gauge configuration. 

Following~\cite{Blum:2001sr}, we define the amputated, bare, vertex function
for each operator $\cal O$,
\begin{equation}
{\cal V^O}(p^2) = \frac{1}{\<S^\dagger(p)\>}\<S^\dagger(p){\cal O}S(p)\>\frac{1}{\<S(p)\>}
\end{equation}
and a corresponding projector that
enforces the tree-level renormalization condition
\begin{eqnarray}
{\rm Tr} {\cal P\, V^O} &\propto & 1. 
\end{eqnarray}
We choose the following,
\begin{eqnarray}
{\cal O}^q_{\{44\}} &\longrightarrow& {{\cal P}^q_{44}}^{-1} =
   \gamma_4 p_4 - \frac{1}{3} \sum_{i=1,3}\gamma_i p_i,\\
{\cal O}^{5q}_{\{34\}} &\longrightarrow& {{\cal P}^{5q}_{34}}^{-1} =
   i\gamma_5\frac{1}{2}[\gamma_3 p_4 + \gamma_4 p_3], \\
{\cal O}^{\sigma q}_{34} &\longrightarrow& {{\cal P}^{\sigma q}_{34}}^{-1} =
   \gamma_5 \sigma_{34}.
\end{eqnarray}
Defining the renormalized operator at scale $\mu^2=p^2$ as ${\cal O}_{ren}(\mu) = Z _{\cal O}(\mu,a){\cal O}(a)$, the renormalized vertex takes the form
\begin{equation}
{\cal V^O}_{ren}(p^2) = \frac{Z_{\cal O}}{Z_q}  {\cal V^O}(p^2),
\end{equation}
and after projection,
\begin{equation}
{\Lambda_{\cal O}}_{ren}(p^2) = 
 \frac{Z_{\cal O}}{Z_q} \, {\Lambda_{\cal O}}(p^2)
=  \frac{1}{Tr{\cal P_O}^2(p)} 
                   Tr\left[{\cal P_O}(p) {\cal V^O}(p^2)\right]\,\,
 \frac{Z_{\cal O}}{Z_q}.
\label{eq:RenLambda}
\end{equation}
From this we extract the required renormalization constant $Z_{\cal O}$ that ensures the same renormalization condition at scale $\mu$ as in the free case.
\begin{equation}
{\Lambda_{\cal O}}_{ren}(\mu^2) = \frac{Z_{\cal O}}{Z_q} \,\, 
{\Lambda_{\cal O}}(\mu^2) =1\,.
\label{eq:RenormCond}
\end{equation}

\subsection{Anomalous dimensions and matching}

In order to compute the renormalization group
invariant (RGI) constants~\cite{Martinelli:1995ty}, we divide out the running of the operator at hand, calculated in continuum perturbation theory.
To be sensible, this is done at large enough momentum 
so the operator runs perturbatively. The momentum
cannot be taken too large, or lattice artifacts will spoil the continuum running of the operator. Thus the momentum scale where the RGI constant is defined must satisfy
\begin{equation}
\Lambda_{\rm QCD} \ll \mu \ll \frac{1}{a}.
\end{equation}

In addition, it is convenient to match the RI/MOM regularization used in NPR with the $\MSbar$ scheme since the latter is conventionally used in the Wilson coefficient calculation.  
For the derivative operators the
matching from the RI/MOM scheme in Landau gauge to $\MSbar$ can be done using the conversion factor~\cite{Gockeler:1999jb}
\begin{equation}
Z^{\MSbar}_{\mathrm {MOM}}(p) = 
    1 + \frac{\alpha_s}{4 \pi} C_F \Bigg[ G_n +  S_{n-1}
  - \frac{2(n-1)}{n(n+1)}
      \frac{\left( \sum_\mu p_\mu h_\mu (p) \right)^2 }
      {p^2 \sum_\mu h_\mu (p)^2}    \Bigg] + O(\alpha_s^2) \,,
\end{equation}
where $C_F=\frac{N_c^2 -1}{2N_c}$ is the quadratic Casimir for the $SU(N_c)$ gauge group, $\alpha_s$ the strong coupling constant,
\begin{eqnarray} 
  G_n & = &  \frac{2}{n(n+1)} \left( 2 S_{n+1} -3-S_{n-1} \right) 
     + \frac{2}{n+1} - 4 \sum_{j=2}^n \frac{1}{j} 
        \left( 2S_j - S_{j-1} \right) -1 \,, \\
  S_n & = & \sum_{j=1}^n \frac{1}{j} \,,
\end{eqnarray}
and
\begin{equation}
 h_\mu (p) = \sum_{\mu_2,\ldots,\mu_n} c_{\mu \mu_2 \ldots \mu_n}
              p_{\mu_2} \cdots p_{\mu_n} \,.
\end{equation}
The conversion factor $Z^{\overline{{MS}}}_{\mathrm{MOM}}$
depends on the direction of the 
momentum $p$ and on the coefficients $c_{\mu_1, \ldots ,\mu_n}$ because the renormalization condition breaks $O(4)$ invariance. The
coefficients $c_{\mu_1, \ldots ,\mu_n}$ are defined using the conventions
in~\cite{Gockeler:1999jb}. In the present case for the momentum fraction and helicity operators we have
\begin{equation}
c_{\mu\nu} = \delta_{\mu4}\delta_{\nu4} -\frac{1}{3}\sum_{k=1}^3\delta_{\mu k}\delta_{\nu k}\,,
\end{equation}
and
\begin{equation}
c_{\mu\nu} = \delta_{\mu3}\delta_{\nu4}  + \delta_{\mu4}\delta_{\nu3}, 
\end{equation}
respectively.

For matching the tensor bilinear we use results 
from~\cite{Gracey:2003yr}.
\begin{equation}
Z^{\overline{{MS}}}_{\mathrm{MOM}} = 1 - \left(\frac{\alpha_s}{4\pi}\right)^2
\frac{C_F}{216}\Bigg[\left(4320\zeta(3) - 4815\right)C_F - 1252 T_F n_f + 
\left(5987 -3024\zeta(3)\right)C_A\Bigg] +O(\alpha_s^3)
\end{equation}
where $\zeta(k)$ is the Riemann zeta function, 
$C_A= N_c$ and $T_F = 1/2$ for the $SU(N_c)$ group.

In our analysis, we first match $\Lambda_{\cal O}$ to $\MSbar$ and then use the $\MSbar$ running of the operators to extract the RGI
renormalization constant.

Following the conventions of~\cite{Blum:2001sr}, the continuum two loop running of the operators is parametrized by
 \begin{equation}
 C_{\cal O}(\mu^2) =
 \alpha_s(\mu)^{\overline{\gamma}_0}
\left\{
1
+ \frac{\alpha_s(\mu)}{4 \pi} \left( \overline{\gamma}_1
-\overline{\beta}_1 \overline{\gamma}_0 \right)
\right\},
 \end{equation}
where the anomalous dimension of the operator is 
\begin{eqnarray}
\gamma_{\cal O}  &=& 
\sum_i \gamma^{(i)}_{\cal O} \left( \frac{\alpha_s}{4 \pi}\right)^{i+1} \\
\overline{\gamma}_{{\cal O}i} &=& \frac{\gamma^{(i)}_{\cal O}}{2 \beta_0} \\
\overline{\beta}_i &=& \frac{\beta_i}{\beta_0} \, ,
\end{eqnarray}
and $\beta_{0,1}$ are the first two coefficients in the weak coupling expansion of the beta function
\begin{equation}
\frac{\beta(\alpha_s)}{4 \pi} = 
- \beta_0 \left[ \frac{ \alpha_s}{ 4\pi } \right]^2
- \beta_1 \left[ \frac{ \alpha_s}{ 4\pi } \right]^3
- \ldots \, .
\end{equation}

The two loop running of $\alpha_s$ is given
by\cite{Buras:1998ra,Gimenez:1998ue}
\begin{equation}
\frac{\alpha_s}{4 \pi}
=
\frac{1}{\beta_0 \ln \left( \mu^2 / \Lambda^2_{QCD} \right) }
-
\frac{\beta_1 \ln \ln \left( \mu^2 / \Lambda^2_{QCD} \right)
}{
\beta_0^3 \ln^2 \left( \mu^2 / \Lambda^2_{QCD} \right)}
\end{equation}
The values of the $\beta_i$'s and the $\gamma_i$'s used in this analysis are given in Tables~\ref{tab:betaFunc}~and~\ref{tab:anomDims} respectively. For $\Lambda_{QCD}$ we take the quenched value~\cite{Capitani:1998mq},
$ \Lambda_{QCD} = 238 \pm 19 \mathrm{MeV}$.

\fi

%%%%%%%%%%%%%%%%%%%%%%%%%%%%  Section %%%%%%%%%%%%%%%%%%%%%%%%%%%%%%%%

\section{Numerical Details}
\label{sec:Numerics}

\ifnum\theNumerics=1
%%%%%%%%%%%%%%%%%%%%%%%%%%%%  Section %%%%%%%%%%%%%%%%%%%%%%%%%%%%%%%%
%
%  \section{Numerical details}
%  \label{sec:Numerics}
%  File:  Numerics.tex
%
%%%%%%%%%%%%%%%%%%%%%%%%%%%%%%%%%%%%%%%%%%%%%%%%%%%%%%%%%%%%%%%%%%%%%%

\ifnum\theOutline=0
\begin{outline}
\item Matrix elemet calculation details
\item lattices methods
\item NPR simulation details
\item Plateaus
\item fits
\item Z factors in MSbar at 2GeV
\end{outline}
\fi

We work in the quenched approximation and use 
domain wall fermions to compute the matrix elements described in the previous sections. We use the DBW2 gauge 
action~\cite{Takaishi:1996xj,deForcrand:1999bi} at
lattice spacing $a^{-1}\approx 1.3$ GeV ($\beta = 0.87$),
with lattice size $16^3\times 32$ and fifth
dimension $L_s = 16$. This action has been shown to significantly reduce the explicit chiral symmetry breaking of domain wall fermions with finite fifth dimension relative to the Wilson gauge action~\cite{Aoki:2002vt}. This relatively coarse lattice spacing was chosen to give a large physical volume ($2.4$ fm spatial
size) and enable calculations with light quark masses to study the chiral behavior of the
matrix elements. An earlier calculation of the nucleon axial charge showed that such a large volume is necessary to avoid significant finite volume errors\cite{Sasaki:2003jh} which, based on that study, we may expect to be a few percent for pion masses in the range $390\MeV$ to $850\MeV$.  The residual quark mass for $L_s=16$ at this lattice spacing and for the DBW2 action is $\mres\approx 0.7$ MeV\cite{Aoki:2002vt}, truly negligible compared to the input quark masses in our simulation, $0.02 \le m_f \le 0.1$, which span a range from about one-quarter to two times the strange quark mass\cite{Aoki:2002vt}.

To calculate three point correlation functions we use box
sources with size $\sim 1.2$ fm and sequential-source propagators
with point sinks. For details see~\cite{Sasaki:2003jh}. The source time is $t=10$ and the source-sink time separation is 10 lattice
units, approximately 1.5 fm, which provides a sufficiently large time separation to observe clear
plateaus. Figures~\ref{fig:plateau-xq}-\ref{fig:plateau-1dq}
show typical plateaus for the matrix elements studied in this work.
Our calculation is done using 416 independent gauge configurations
produced using the overelaxed heatbath algorithm described
in~\cite{Aoki:2002vt}. All statistical errors are
jackknife estimates.

For the calculation of renormalization constants we use 120 lattices and fix
to Landau gauge using the technique described in~\cite{Blum:2001sr}.
Two loop continuum running is used to extract the RGI renormalization
constants in all cases. In order to eliminate remaining scaling violations we fit the data linearly in $(ap)^2$ and define the renormalization constant from the intercept 
(as done in~\cite{Blum:2001sr}). The fitting range used
is $(pa)^2 \in [1.2,1.9]$.

\fi

%%%%%%%%%%%%%%%%%%%%%%%%%%%%  Section %%%%%%%%%%%%%%%%%%%%%%%%%%%%%%%%

%%%%%%%%%%%%%%%%%%%%%%%%%%%%  Section %%%%%%%%%%%%%%%%%%%%%%%%%%%%%%%%

%\section{Renormalization Constants}
%\label{sec:LatticeRenorm}

%\ifnum\theLatticeRenorm=0
%\input{text_sections/LatticeRenorm.tex}
%\fi

%%%%%%%%%%%%%%%%%%%%%%%%%%%%  Section %%%%%%%%%%%%%%%%%%%%%%%%%%%%%%%%

\section{Results and Discussion}
\label{sec:Results}

\ifnum\theResults=1
%%%%%%%%%%%%%%%%%%%%%%%%%%%%  Section %%%%%%%%%%%%%%%%%%%%%%%%%%%%%%%%
%
%  \section{Results and Discussion}
%  \label{sec:Results}
%  File:  Results.tex
%
%%%%%%%%%%%%%%%%%%%%%%%%%%%%%%%%%%%%%%%%%%%%%%%%%%%%%%%%%%%%%%%%%%%%%%

\ifnum\theOutline=0
\begin{outline}
\item Renormalized matrix elements at 2GeV MSbar
\item Discussion of the results
\end{outline}
\fi

The nucleon mass has been determined on a subset of this ensemble of
lattices previously\cite{Aoki:2002vt}. For the present analysis, we
fit the two-point correlation function to a single exponential from
time $t=6$ to 15. The fitted nucleon mass for each quark mass is given
in Table~\ref{tab:nucleon mass}; they are consistent with those
reported in \cite{Aoki:2002vt}.

The central question we sought to answer with this study is why
lattice results for the first moments of the polarized and unpolarized
structure functions disagree with fits to experimental
measurements\cite{Gockeler:1996wg,Dolgov:2002zm,
Detmold:2001jb,Arndt:2001ye,Chen:2001gr,Gockeler:2004wp}.  Preliminary
results from this study and one using two flavors of dynamical domain
wall fermions have been reported in~\cite{Orginos:2002mn,Ohta:2003ux,Ohta:2004mg}.
The discrepancy is large ($\simge50$\%), and holds for dynamical as
well as quenched calculations. A plausible explanation is that the
quarks simulated in these past studies, valence and sea, have been too
heavy.  Here we use as light a quark mass that has been simulated to
date for nucleon structure calculations, roughly one-quarter of the
strange quark mass, in order to investigate the chiral regime. If the
problem is related to the sea quark mass, a resolution will have to
wait for future dynamical fermion calculations using lighter sea quark
masses than have been used already\footnote{The RBC and UKQCD collaborations are embarking on a large scale project
this year to generate an extensive ensemble of 2+1 flavor domain wall
fermion lattices with sea quark masses as light as 1/5 the strange
quark mass. Among many other quantities, nucleon structure will be
studied}.

In Figure~\ref{fig:Xq} and Table~\ref{tab:unpol} we present our
results for the momentum fraction, $\<x\>_q$, of the up and down
valence quarks. These results do not contain disconnected
diagrams. Figure~\ref{fig:Xq_ns} shows the isovector matrix element
$\<x\>_{u-d}$ where disconnected diagrams do not contribute (for
degenerate u, d quarks). The central values are obtained from a
constant fit over the range $13 \le t \le 16$, based on the plateau in
Figure~\ref{fig:plateau-xq} (the same range is used for all matrix
elements in this work).  The quark mass dependence is mild and appears
linear.  The renormalization constant corresponding to the momentum
fraction is $Z^{\MSbar}(2$~GeV)$=1.02(10)$, essentially one, (see
Figure~\ref{fig:NPR-Xq} and Table~\ref{tab:Zfact}). The renormalized
results are all significantly higher than the value $0.154(3)$ 
extracted from the experimental data 
of~\cite{Lai:1996mg,Gluck:1998xa,Martin:2001es} by Dolgov
et.al.~\cite{Dolgov:2002zm}, and there is no apparent curvature as
$m_\pi^2\to 0$ that leads us to believe that in the chiral limit the
two results would agree. Thus, we do not attempt to extrapolate to
$m_\pi^2=0$. This is the same behavior witnessed in previous studies;
in particular, our $\overline{MS}$ values for $\<x\>_q$ are quite
consistent with the recently reported quenched improved Wilson
fermion, continuum and chiral limit, value in~\cite{Gockeler:2004wp},
 suggesting good scaling of domain wall fermions.

Figure~\ref{fig:XDq} shows the first moment of the helicity
distribution $\<x\>_{\Delta q}$ for the up and down quarks (tabulated
in Table~\ref{tab:pol}). As before, these results do not contain
disconnected diagrams, and in Figure~\ref{fig:XDq_ns} we display the
isovector matrix element $\<x\>_{\Delta u - \Delta d}$ where
disconnected diagrams do not contribute.  Again, the quark mass
dependence is mild and appears linear, the renormalization constant is
essentially unity, $Z^{\MSbar}(2$~GeV$)=1.02(9)$ (see
Figure~\ref{fig:NPR-XDq} and Table~\ref{tab:Zfact}), and the
renormalized moments lie above the value $0.196(4)$ extracted from the
experimental data of~\cite{Gluck:1995yr,Gehrmann:1995ag} by Dolgov
et.al.~\cite{Dolgov:2002zm}.

Since chiral symmetry requires that the renormalization constants of
the momentum fraction and the helicity distribution be the same, we
can consider the ratio of the bare matrix elements in which the
renormalization constants and matching factors cancel.  A similar
ratio worked well in the case of the axial charge,
$g_A$ \cite{Sasaki:2003jh}, and we saw already that the explicit
calculations of these constants gave the same result well within
statistical errors.  Figure~\ref{fig:Xq_o_XDq} shows the ratio
together with the value extracted from
experiment\cite{Gluck:1995yr,Gehrmann:1995ag,Lai:1996mg,Adams:1997tq,Adeva:1997qz,Gluck:1998xa,Ackerstaff:1999ey,Martin:2001es,Dolgov:2002zm}. As it is argued 
in~\cite{Dolgov:2002zm}, it is difficult to estimate the systematic errors
associated with the experimental extraction of the values of both
$\<x\>_{u-d}$ and $\<x\>_{\Delta u - \Delta d}$ but it is almost certain
that these systematic errors are smaller than the statistical
errors of the lattice calculation. For that reason the comparison
of lattice results to experiment is not out of order.
Interestingly, our results are in very good agreement with experiment; a discrepancy is not evident at all. Note that the jackknife determination of the ratio is determined relatively precisely since the error on the renormalization constants does not enter and the numerator and denominator are highly correlated.

In \cite{Sasaki:2003jh} the axial and vector matrix elements displayed
very different finite volume behavior as the quark mass was reduced;
the axial matrix element, being sensitive to low energy physics,
decreased drastically in the limit $m_f\to 0$ in the small volume
study. If similar behavior is operative here, the mild quark mass
dependence of the matrix elements suggests that finite volume effects
are small.

%Chen:2001eg full qcd result
In \cite{Chen:2001gr} the quark mass dependence near the chiral limit
of $\<x\>_{u-d}$ but not of $\<x\>_{\Delta u-\Delta d}$, was computed
in quenched chiral perturbation theory, so we can not say what the
chiral perturbation theory prediction is for
the ratio.  For $\<x\>_{u-d}$ alone, the chiral perturbation theory
formula allows a wide range of values at the physical point, depending
on the values of various unknown low energy constants.  The authors
of \cite{Chen:2001gr} investigate the quark mass dependence for
several values of the unknown parameters. The most natural choice
demonstrates a smaller dependence on the quark mass than that
predicted by full QCD chiral perturbation theory.

Our results suggest that whatever systematic error causes the
discrepancy in the individual moments from experimental expectations
appears to mostly cancel in their ratio. Although our calculation is quenched,
it is instructive to look at the full QCD chiral perturbation theory formulas found in~\cite{Chen:2001eg,Arndt:2001ye}. 
\begin{eqnarray}
\<x\>_{u-d} &=& C \left[ 1 - \frac{3g_A^2 + 1}{(4\pi f_\pi)^2} m_\pi^2 \ln\left(\frac{m_\pi^2}{\mu^2}\right) +e(\mu^2)\frac{m_\pi^2}{(4\pi f_\pi)^2}\right] \\
\<x\>_{\Delta u- \Delta d} &=& \tilde{C} \left[ 1 - \frac{2g_A^2 + 1}{(4\pi f_\pi)^2} m_\pi^2 \ln\left(\frac{m_\pi^2}{\mu^2}\right) + \tilde{e}(\mu^2)\frac{m_\pi^2}{(4\pi f_\pi)^2}\right] 
\end{eqnarray}
where the normalization is such that the physical pion decay constant
is $f_\pi=93$ MeV, $C$ and $\tilde C$ are unknown constants, and
$e(\mu^2)$ and $\tilde e(\mu^2)$ are counter terms evaluated at the
renormalization scale $\mu$.  In Figure~\ref{fig:ChiPT} we plot the
above formulas for $\mu = 1$ GeV. The unknown constants $C$ and
$\tilde C$ are chosen so that the formulae reproduce the experimental
result at the physical pion mass point, and the counter terms are set
to zero in order to isolate the effect of the chiral logarithm.  We
see that there is a strong dependence on the pion mass. As a result,
when $m_\pi\approx400$ MeV, the momentum fraction is roughly $50\%$
larger than at the physical point, while the first moment of the
helicity is about $30\%$ larger. This indicates that the discrepancy
between the lattice data and experiment may be due to the unphysically
large masses used in current lattice simulations. On the other hand,
the large size of the one-loop perturbative corrections suggests that
chiral perturbation theory at this order is unreliable. The effects of
the counter terms and higher order contributions may also be large and
could tend to cancel the large one-loop corrections.  Nevertheless,
it is worth noting that our
quenched results at pion masses of about 400 MeV differ from
experiment by roughly the same amount as one-loop chiral perturbation theory in
full QCD predicts.  The observation of the large one-loop corrections
was first made by the authors
of~\cite{Detmold:2001jb,Detmold:2002ac,Detmold:2002nf,Thomas:2002sj}.

On the other hand the ratio of the momentum fraction to the first
moment of the helicity (Figure~\ref{fig:ChiRatio}) is a milder
function of the quark mass. The difference of the experimental result
from the value at $m_\pi=400$ MeV is about $10\%$, not wildly
inconsistent with the lattice result shown in
Figure~\ref{fig:Xq_o_XDq}.

The conclusion from this discussion is that the discrepancy between lattice and experimental results for the quark momentum fraction and helicity distribution is most likely due to strong mass dependence in these functions and will be resolved by pushing lattice simulations further into the light quark mass region. Since dynamical results so far are similar to quenched\cite{Ohta:2003ux,Ohta:2004mg}, this probably holds in that case too.
It may also be of interest to obtain the next higher order results in chiral perturbation theory for these quantities, though this seems a more daunting task.

We now turn to another interesting feature in our calculation, the
twist-3 matrix element $d_1$.  Although it is not measurable in deep
inelastic scattering of electrons on protons, it serves as an example
of what can be expected for the $d_n$ matrix elements.  As discussed
in Section~\ref{sec:Renorm}, the operator used to calculate $d_1$,
${\cal O}^{5q}_{[\mu\nu]}$, mixes with the lower dimensional operator
${\cal O}^{\sigma q}_{\mu\nu}$ when chiral symmetry is explicitly
broken.  With domain wall fermions, unlike Wilson fermions, chiral
symmetry is not broken ($m_{res}$ is small enough to ignore in this
study), so power divergent mixing should not occur. Results are
summarized in Table~\ref{tab:d1v}. Figure~\ref{fig:d1} shows, unlike
the Wilson fermion result~\cite{Dolgov:2002zm}, this matrix element is
small in the chiral limit. In fact, the power divergence in the Wilson
fermion case switches the sign of the u and d quark
contributions. Using Wilson fermions, the QCDSF collaboration found
similar results for $d_2$ after a non-perturbative subtraction of the
power divergence\cite{Gockeler:2000ja}.  These results confirm our expectations that the
domain wall fermion formulation avoids the power divergence present
for Wilson fermions.  Because the value of $d_1$ computed here and
QCDSF's value of $d_2$ appear small in the chiral limit, we conclude
that the Wandzura-Wilczek relation between moments of $g_1$ and
$g_2$~\cite{Wandzura:1977qf}, which assumes vanishing $d_n$, is at
least approximately true. This relation is not obvious in a confining
theory~\cite{Jaffe:1991qh}.

Finally, we have computed the first moment of the transversity
distribution, $\<1\>_{\delta q}$, for up and down quarks
(Figure~\ref{fig:1dq}) and the isovector combination $\<1\>_{\delta u
- \delta d}$ (Figure~\ref{fig:1dq_ns}).  The transversity is an
important target of the RHIC spin program at Brookhaven National
Laboratory (see~\cite{Surrow:2002zr,Bland:2004wu} and references
therein).  In both cases the quark mass dependence is mild and appears
to be linear.  The renormalization constant (Figure~\ref{fig:NPR-1dq}
and Table~\ref{tab:Zfact}) is $Z^{\MSbar}(2$ GeV$)=0.872(11)$. Naive
linear extrapolation to the chiral limit yields $\<1\>_{\delta u
- \delta d} = 1.193(30)$ for $\mu= 2$ GeV in the $\overline{MS}$
scheme.  This result contains an unknown systematic error from the
chiral extrapolation which may be small given the mass dependence
exhibited in Figure~\ref{fig:1dq_ns}. We note that the calculation of
the flavor non-singlet tensor charge is similar to $g_A$ which also
exhibited mild mass dependence and whose value agrees well with
experiment\cite{Sasaki:2003jh}.  We caution the reader that the
momentum fraction and helicity distribution also exhibited mild mass
dependence but are known to disagree with fits to the experimental
data.  Our value is consistent with another recent
calculation~\cite{Gockeler:2005aw}.

\fi

%%%%%%%%%%%%%%%%%%%%%%%%%%%%  Section %%%%%%%%%%%%%%%%%%%%%%%%%%%%%%%%

\section{Conclusions}
\label{sec:conclusions}

\ifnum\theConclusions=1
%%%%%%%%%%%%%%%%%%%%%%%%%%%%  Section %%%%%%%%%%%%%%%%%%%%%%%%%%%%%%%%
%
%  \section{Conclusions}
%  \label{sec:conclusions}
%  File:  conclusions.tex
%
%%%%%%%%%%%%%%%%%%%%%%%%%%%%%%%%%%%%%%%%%%%%%%%%%%%%%%%%%%%%%%%%%%%%%%

\ifnum\theOutline=0
\begin{outline}
\item Don't know yet....
\end{outline}
\fi

We have reported on a quenched calculation of the first moments of the
polarized and unpolarized structure functions of the nucleon.  We have
used domain wall fermions in a relatively large physical volume, $\sim
(2.4$ fm)$^3$.  The large volume is important to minimize finite
volume errors in nucleon matrix element calculations as shown earlier
in the calculation of the axial charge, $g_A$ \cite{Sasaki:2003jh}.
The chiral symmetry of domain wall fermions may also be useful in
unraveling the mystery of the large discrepancy between lattice
calculations and experiment for the moments of structure functions.
While our results for the individual moments show no evidence of the
chiral logarithm predicted
in~\cite{Chen:2001gr,Arndt:2001ye,Chen:2001eg,Chen:2001nb} for pion
masses as low as $390$ MeV, the ratio of the momentum fraction to the
first moment of the helicity distribution is in very good agreement
with the experimental value (Figure~\ref{fig:Xq_o_XDq}). We note that
the ratio is computed on the lattice more accurately than the
individual moments. This agreement, taken together with chiral
perturbation theory calculations, leads us to conclude that the
discrepancy between lattice calculations and fits to experiment is
most likely due to strong mass dependence of these functions in the
light quark regime. Thus, ultimately the difference will be resolved
as lattice calculations push into this regime.

A recent large scale, detailed, investigation of the discrepancy in
low moments of nucleon structure functions was reported
in \cite{Gockeler:2004wp}; using improved Wilson fermions, the authors
performed continuum and chiral extrapolations but did not resolve the
problem. The agreement with our results suggests that scaling
violations for domain wall fermions are mild (the coarsest lattice
spacing used in ~\cite{Gockeler:2004wp} was about 0.1 fm compared to
0.15 fm used here).

Our calculation of the $d_1$ matrix element (Figure~\ref{fig:d1})
indicates that the domain wall formalism eliminates power divergent
mixing in this class of matrix elements. This suggests that a
precision calculation of $d_2$ with domain wall (or other chiral)
fermions is possible and should be undertaken. Since the mixing with
lower dimensional operators induced by explicit chiral symmetry
breaking is linearly divergent with $a^{-1}$, care must be taken to
minimize $m_{res}$.

Finally, we have computed the first moment of the transversity
distribution, or tensor charge, which will be measured at Brookhaven
National Laboratory as part of the RHIC spin program.  We find
$\<1\>_{\delta u - \delta d} = 1.193(30)$ at $\mu=2$ GeV in the
$\overline{MS}$ scheme, with unknown systematic error stemming from
the linear chiral extrapolation. The mild quark mass dependence in the
tensor charge suggests this systematic error is small, as does the
similarity to the calculation of $g_A$. On the other hand, such mild
mass dependence is also observed in the momentum fraction and helicity
distribution, hence further study of the chiral extrapolation of this
observable is needed in order to obtain a reliable estimate of the systematic
error. 

An unknown systematic error due to quenching exists in our
results. Two flavor calculations using Wilson fermions with relatively
heavy quark masses exist
\cite{Dolgov:2002zm,Gockeler:2004vx}, we have begun two flavor 
domain wall fermion
calculations \cite{Ohta:2003ux,Ohta:2004mg,Aoki:2004ht}, and 2+1
flavor domain wall fermion calculations are just beginning.  These
studies begin to address the quenching error.  In addition, systematic
uncertainties due to continuum, chiral and infinite volume
extrapolations must be addressed as in the extensive quenched study
using improved Wilson fermions reported in~\cite{Gockeler:2004wp}.
This study suggests domain wall fermions should facilitate these
extrapolations. Recent results from chiral perturbation
theory~\cite{Beane:2003xv,Detmold:2005pt} may also prove useful.

\fi

%%%%%%%%%%%%%%%%%%%%%%%%%%%%  Section %%%%%%%%%%%%%%%%%%%%%%%%%%%%%%%%

\ifnum\theAcknowledgments=1
\begin{acknowledgments}

We thank our colleagues in the RBC collaboration, Y. Aoki, C. Dawson, N. Christ, T. Izubuchi, C. Jung, R. Mawhinney,  and A. Soni for useful and stimulating discussions. KO also thanks M. Savage for helpful suggestions on the text.

The numerical computations reported here were done on the 400 Gflops QCDSP
supercomputer \cite{Chen:1998cg} at Columbia University and the 600 Gflops QCDSP supercomputer \cite{Mawhinney:2000fx} at the RIKEN BNL Research Center. We thank RIKEN, Brookhaven National Laboratory and the U.S. Department of Energy for providing the facilities essential for the completion of this work.

This research was supported in part by the RIKEN BNL Research
Center and in part by DOE grant DF-FC02-94ER40818 (Orginos).

\end{acknowledgments}
\fi

%%%%%%%%%%%%%%%%%%%%%%%%%%%%%% APPENDIX %%%%%%%%%%%%%%%%%%%%%%%%%%

\appendix

\ifnum\theAppendix=1
%%%%%%%%%%%%%%%%%%%%%%%%%%%%  Section %%%%%%%%%%%%%%%%%%%%%%%%%%%%%%%%
%
%  File:  appendix.tex
%
%%%%%%%%%%%%%%%%%%%%%%%%%%%%%%%%%%%%%%%%%%%%%%%%%%%%%%%%%%%%%%%%%%%%%%

\ifnum\theOutline=1
\begin{outline}
\item
\end{outline}
\fi

%%%%%%%%%%%%%%%%%%%%%%%%%%  Section  %%%%%%%%%%%%%%%%%%%%%%%%%%%%%%%%%

\section{There is no apendix}

\fi

%%%%%%%%%%%%%%%%%%%%%%%%%%%%  Bibliography %%%%%%%%%%%%%%%%%%%%%%%%%%%

\bibliography{paper}

\pagebreak

%%%%%%%%%%%%%%%%%%%%%%%%%%%%%% TABLES %%%%%%%%%%%%%%%%%%%%%%%%%%

\ifnum\theTables=1

\begin{table}[!htb]
\caption{The one- and two-loop expansion coefficients of the quenched QCD $\beta$ function.}
\label{tab:betaFunc}
\begin{ruledtabular}
\begin{tabular}{cc}
$\beta_i$ & quenched ($n_f=0$) value \\ \hline
$\beta_0$ & 11    \\ 
$\beta_1$ & 102   \\ 
%$\beta_2$ & 1428.5\\ 
\end{tabular}
\end{ruledtabular}
\end{table}
\begin{table}[!htb]
\caption{Anomalous dimensions of quark bilinear operators used in this paper in quenched ($n_f=0$) QCD. }
\label{tab:anomDims}
\begin{ruledtabular}
\begin{tabular}{ccc}
Operator & $\gamma_0$ & $\gamma_1$ \\ \hline
${\cal O}^q_{44}$         & 64/9 & 96.69 \\
${\cal O}^{5q}_{\{34\}}$   & 64/9 & 96.69 \\
${\cal O}^{\sigma q}_{34}$ &  8/3 & 724/9 \\
\end{tabular}
\end{ruledtabular}
\end{table}

\begin{table}[htb]
\caption{The nucleon mass in lattice units. Values are from fully covariant, single exponential, fits to the two-point correlation functions in the range $6\le t \le 15$. Errors are statistical only.}
\label{tab:nucleon mass}
\newcommand{\m}{\hphantom{$-$}}
\newcommand{\cc}[1]{\multicolumn{1}{c}{#1}}
\begin{ruledtabular}
\begin{tabular}{lll}
$m_f$        & $m_N$ (error) & $\chi^2$ (dof)  \\
\hline
0.02 & 0.856 (10) & 5.9 (8) \\
0.04 & 0.967 (6) & 1.9 (8)\\
0.06 & 1.064 (5)& 3.5 (8) \\
0.08 & 1.155 (4) & 5.1 (8)\\
0.10 & 1.241 (4)& 6.4 (8)\\
\end{tabular}\\
\end{ruledtabular}
\end{table}

\begin{table}[htb]
\caption{Lowest moment of unpolarized structure functions for u and d quarks and the flavor non-singlet combination. Errors are statistical only. Values are not renormalized.}
\label{tab:unpol}
\newcommand{\m}{\hphantom{$-$}}
\newcommand{\cc}[1]{\multicolumn{1}{c}{#1}}
\begin{ruledtabular}
\begin{tabular}{llll}
$m_f$        & $\<x\>_{u}$   & $\<x\>_{d}$  & $\<x\>_{u-d}$  \\
\hline
0.020 &    0.446(18) &    0.184(11) &    0.262(16)\\ 
0.040 &    0.446(7) &    0.191(4) &    0.255(6)\\ 
0.060 &    0.457(5) &   0.2012(28) &    0.256(3)\\ 
0.080 &    0.464(4) &   0.2092(21) &   0.2548(25)\\ 
0.100 &    0.471(3) &   0.2160(18) &   0.2549(21)\\ 
\end{tabular}\\
\end{ruledtabular}
\end{table}

\begin{table}[htb]
\caption{Lowest moment of the polarized structure fucntions. Errors are statistical only. u and d valence quark contributions. $\Delta u - \Delta d$ denotes the flavor non-singlet combination. Errors are statistical only. Values are not renormalized.}
\label{tab:pol}
\newcommand{\m}{\hphantom{$-$}}
\newcommand{\cc}[1]{\multicolumn{1}{c}{#1}}
\begin{ruledtabular}
\begin{tabular}{llll}
$m_f$        &  $\<x\>_{\Delta u}$   & $\<x\>_{\Delta d}$  & $\<x\>_{\Delta u-\Delta d}$ \\
\hline
0.020 &    0.271(21) &   -0.060(13) &    0.331(23) \\ 
0.040 &    0.261(7) &   -0.069(4) &    0.330(8) \\ 
0.060 &    0.260(4) &  -0.0671(23) &    0.328(5) \\ 
0.080 &    0.263(3) &  -0.0664(16) &    0.330(4) \\ 
0.100 &   0.2667(26) &  -0.0667(12) &   0.3335(29)\\ 
\end{tabular}\\
\end{ruledtabular}
\end{table}

\begin{table}[htb]
\caption{The twist-3 matrix element $d_1$. u and d valence quark contributions. Errors are statistical only. Values are not renormalized.}
\label{tab:d1v}
\newcommand{\m}{\hphantom{$-$}}
\newcommand{\cc}[1]{\multicolumn{1}{c}{#1}}
\begin{ruledtabular}
\begin{tabular}{lll}
$m_f$       &$d_{1u}$  & $d_{1 d}$  \\
\hline
0.020 &    0.045(21) &   -0.029(14)\\ 
0.040 &    0.075(8) &   -0.031(5)\\ 
0.060 &    0.109(5) &  -0.0354(29)\\ 
0.080 &    0.140(4) &  -0.0405(20)\\ 
0.100 &    0.167(3) &  -0.0457(15)\\ 
\end{tabular}\\
\end{ruledtabular}
\end{table}

\begin{table}[htb]
\caption{The  $\MSbar$ renormalization constants at $\mu=2$ GeV. Errors are statistical only.}
\label{tab:Zfact}
\begin{ruledtabular}
\begin{tabular}{ll}
$Z_{\<x\>_q}$         &  1.02(9) \\  
$Z_{\<x\>_{\Delta q}}$ &  1.02(10) \\
$Z_T$                &  0.872(11) \\  
\end{tabular}\\
\end{ruledtabular}
\end{table}

\clearpage
\pagebreak

\fi
\pagebreak

%%%%%%%%%%%%%%%%%%%%%%%%%%%%%% FIGURES %%%%%%%%%%%%%%%%%%%%%%%%%%%%%%%%

\ifnum\theFigures=1
%%%%%%%%%%%%%%%%%%%%%%%%%%%%%%%%%%%%%%%%% 
%
%  Figures 
%
\begin{figure}
\begin{center}
\includegraphics[width=\textwidth]{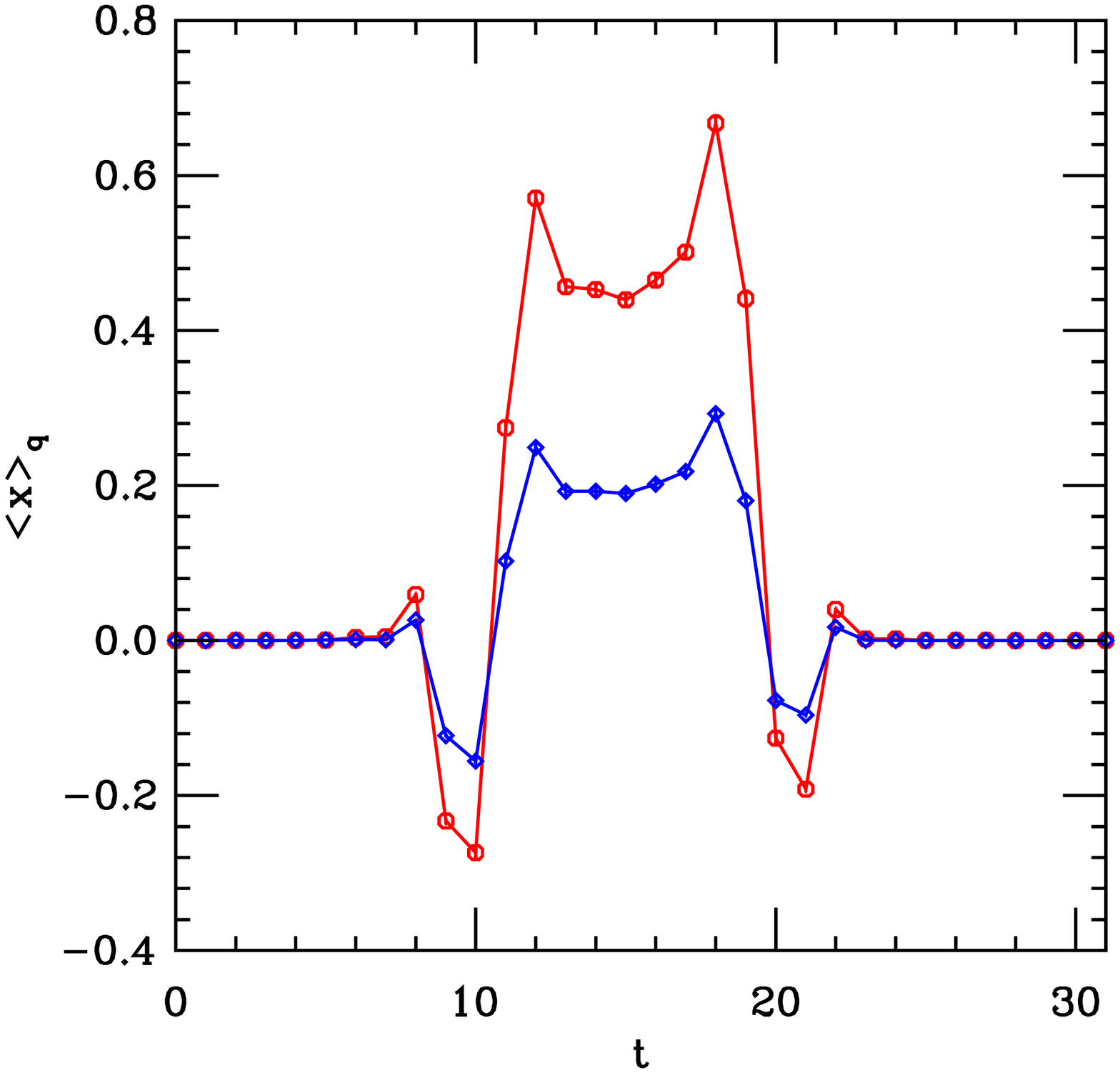}
\end{center}
\caption{The three point correlation function for the operator ${\cal O}^q_{44}$ and bare quark mass $m_f=0.04$. Octagons are the up quark contribution and diamonds are the down quark contribution.}
\label{fig:plateau-xq}
\end{figure}

\begin{figure}
\begin{center}
\includegraphics[width=\textwidth]{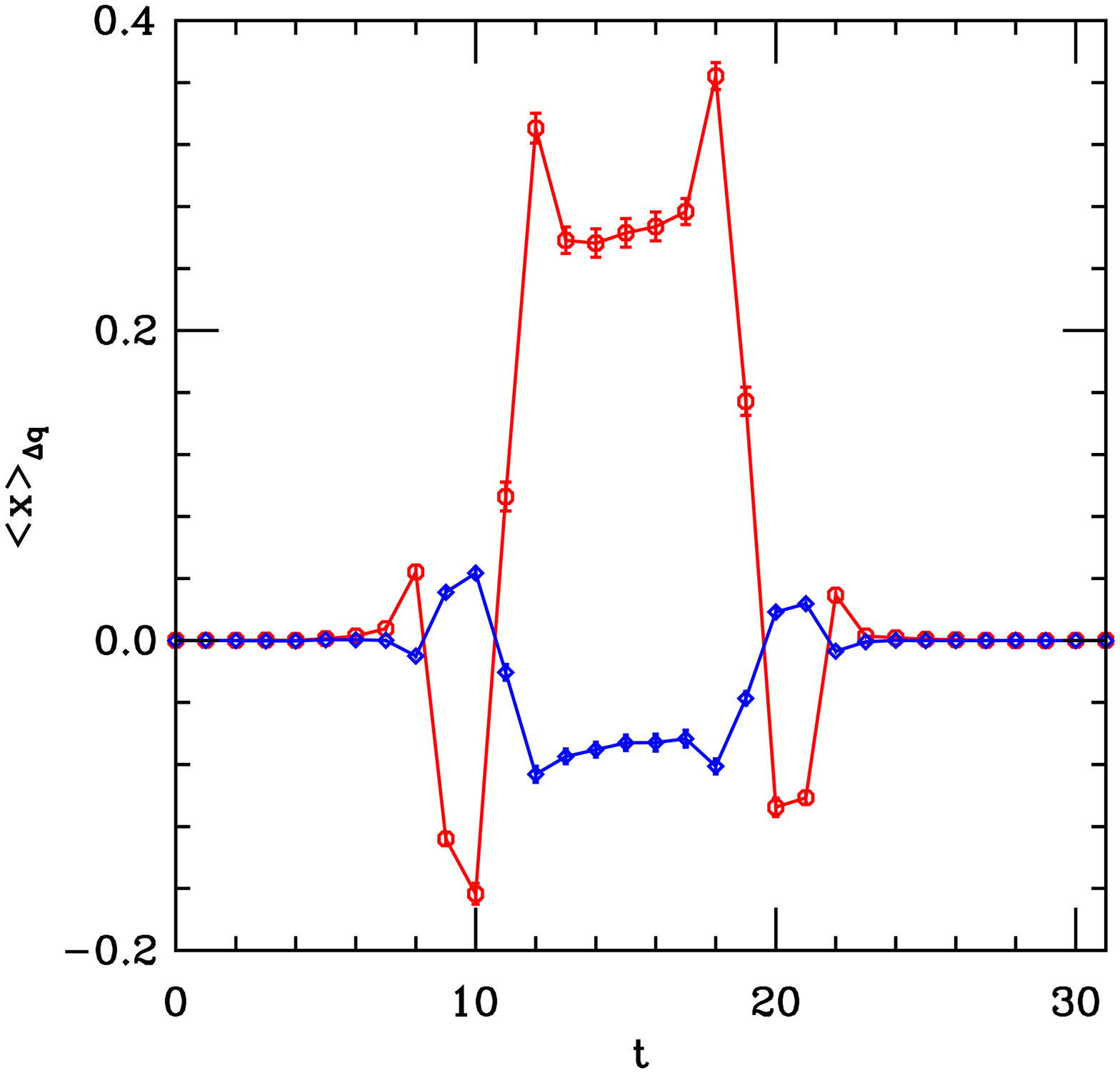}
\end{center}
\caption{The three point correlation function for the operator ${\cal O}^{5q}_{\{34\}}$ and bare quark mass 0.04. Octagons are the up quark contribution and diamonds are the down quark contribution.}
\label{fig:plateau-xDq}
\end{figure}

\begin{figure}
\begin{center}
\includegraphics[width=\textwidth]{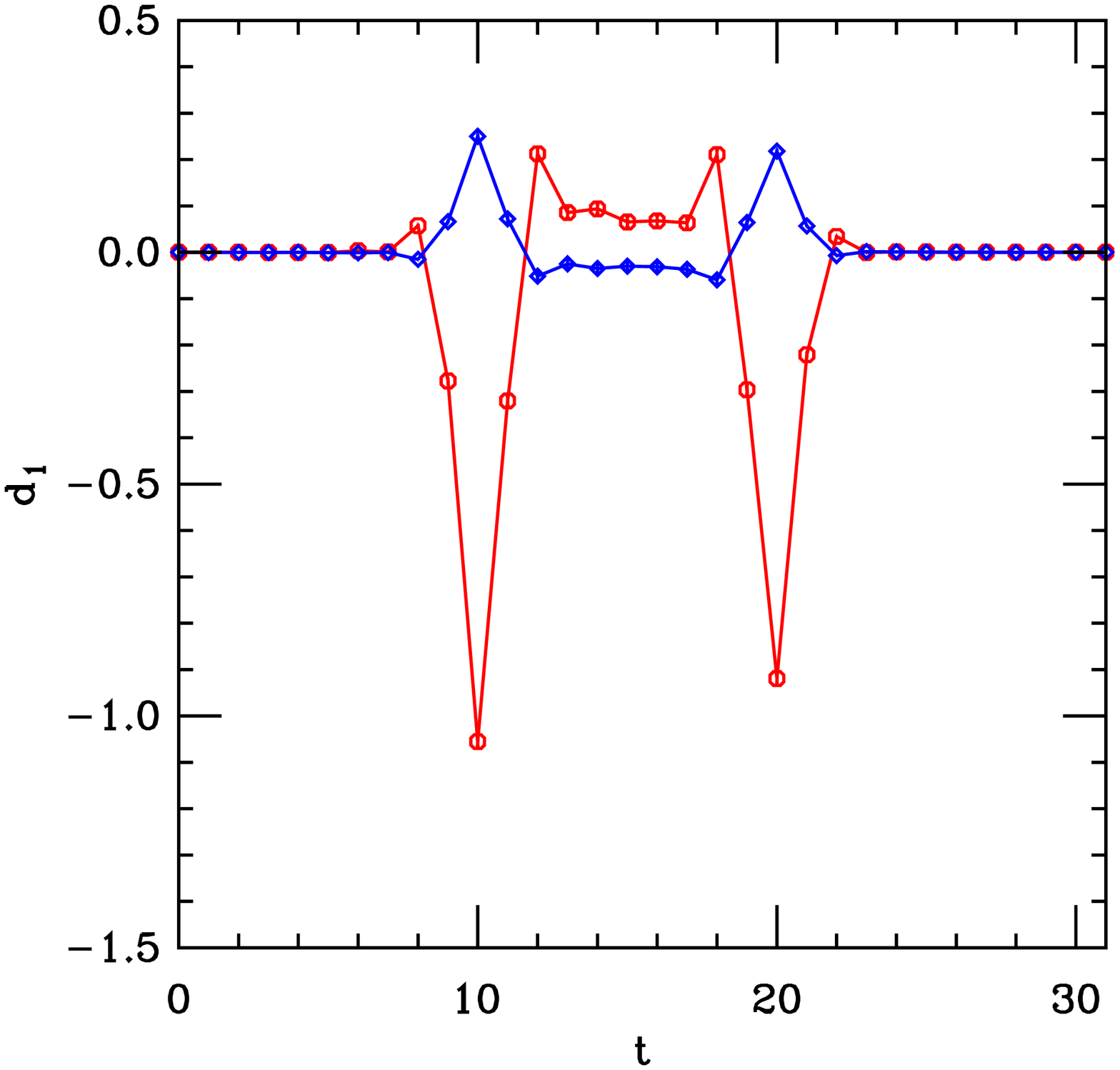}
\end{center}
\caption{The three point correlation function for the operator ${\cal O}^{5q}_{[34]}$ and bare quark mass 0.040. Octagons are the up quark contribution and diamonds are the down quark contribution.}
\label{fig:plateau-d1}
\end{figure}

\begin{figure}
\begin{center}
\includegraphics[width=\textwidth]{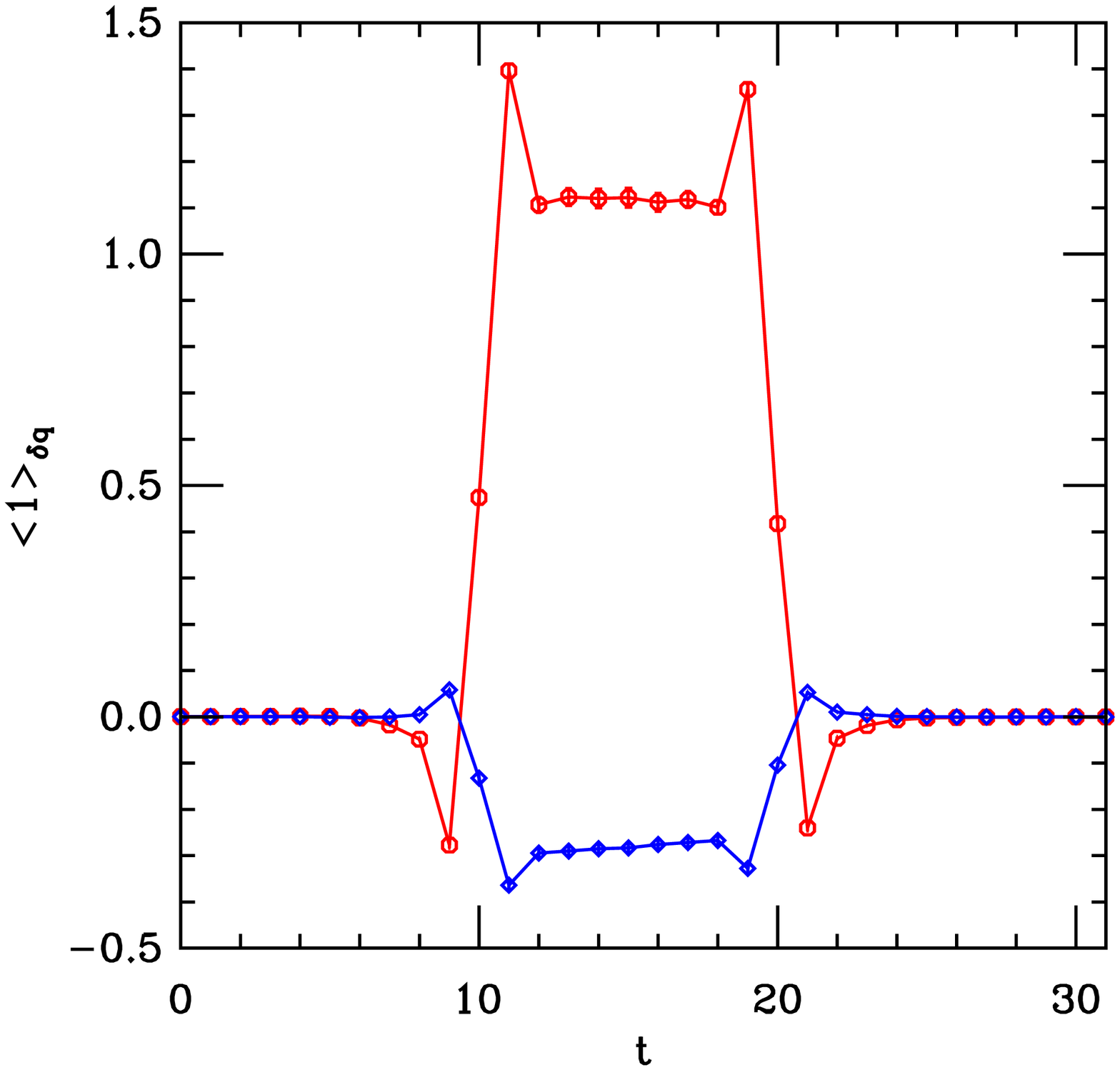}
\end{center}
\caption{The three point correlation function for the operator ${\cal O}^{\sigma q}_{34}$ and bare quark mass 0.04. Octagons are the up quark contribution and diamonds are the down quark contribution.}
\label{fig:plateau-1dq}
\end{figure}

\begin{figure}
\begin{center}
\includegraphics[width=\textwidth]{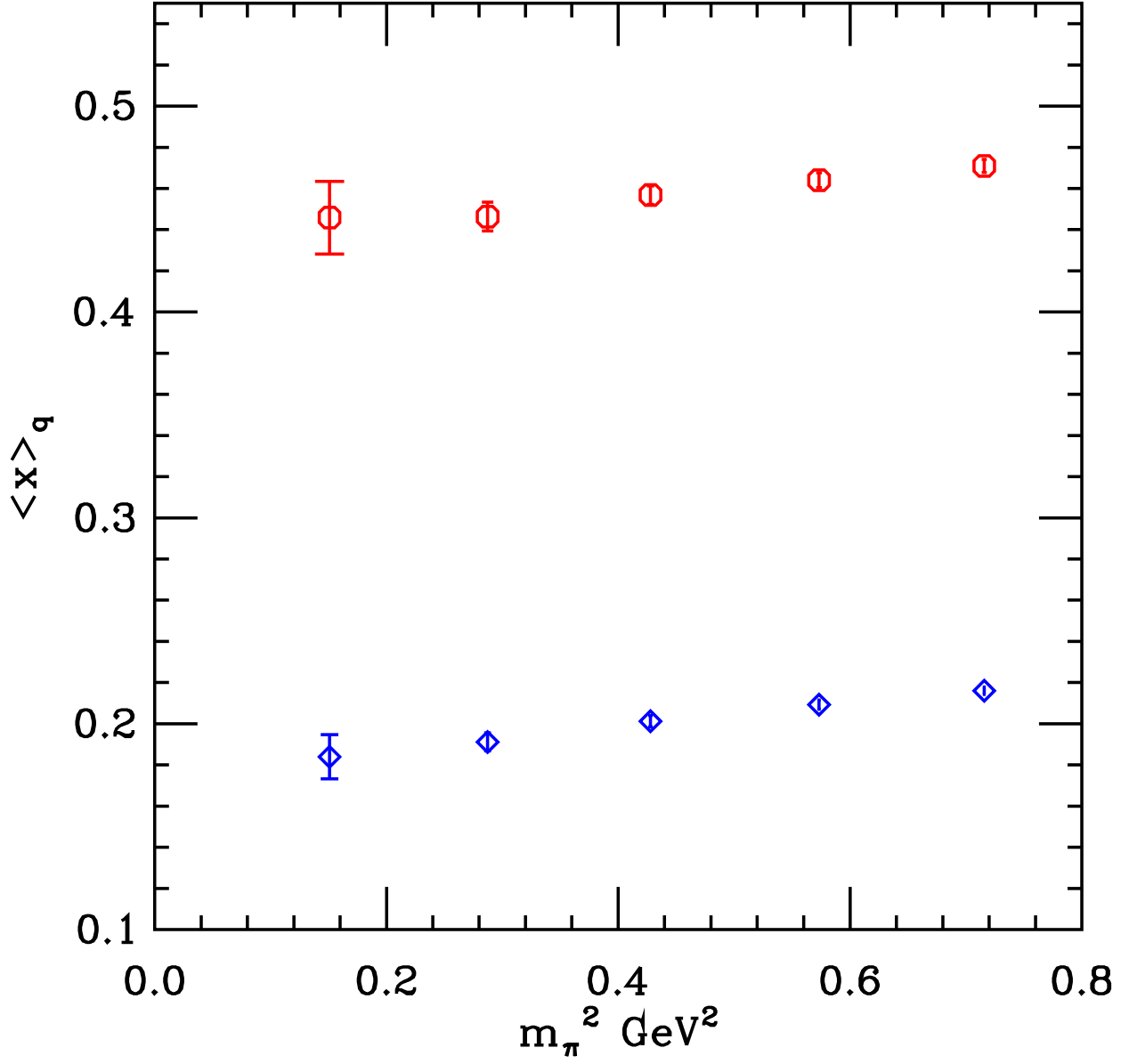}
\end{center}
\caption{The bare momentum fraction. $m_\pi$ is the pseudoscalar mass. Octagons are the up quark contributions and diamonds are the down quark contributions. Disconnected diagrams are not included.}
\label{fig:Xq}
\end{figure}

\begin{figure}
\begin{center}
\includegraphics[width=\textwidth]{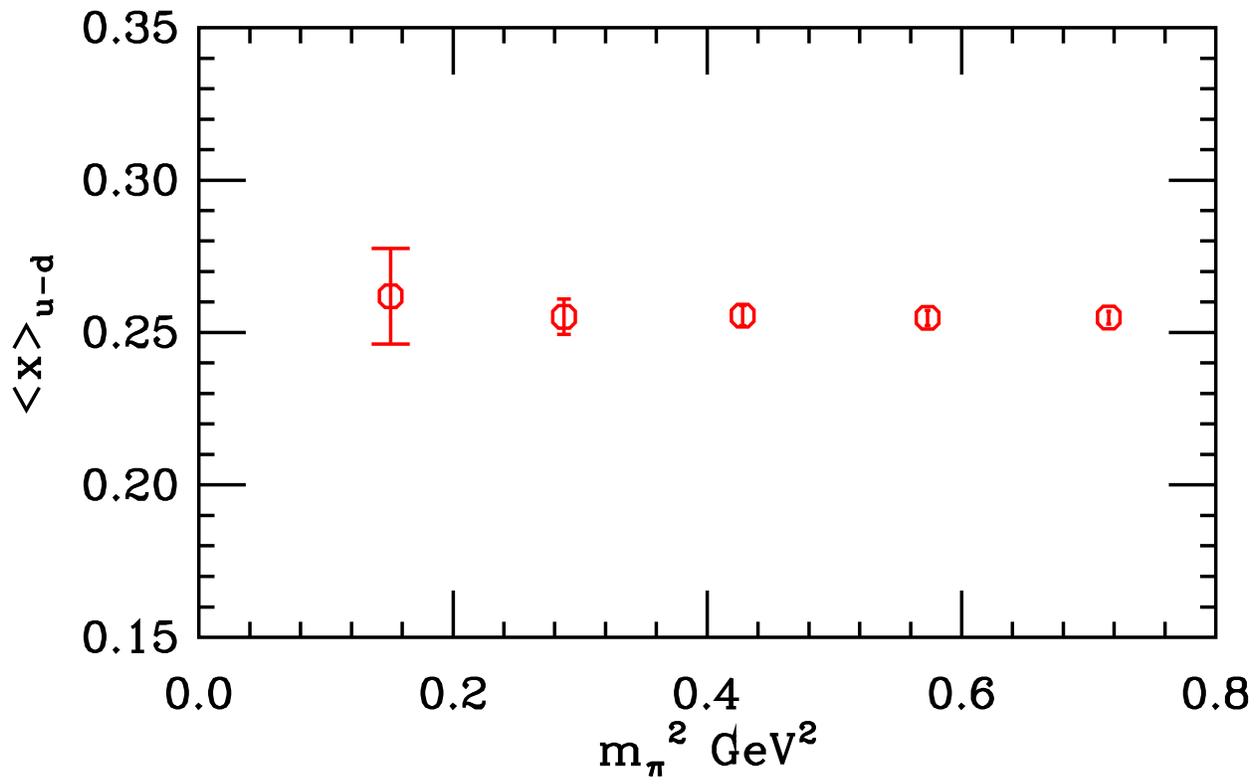}
\end{center}
\caption{The bare flavor non-singlet momentum fraction.}
\label{fig:Xq_ns}
\end{figure}

\begin{figure}
\begin{center}
\includegraphics[width=\textwidth]{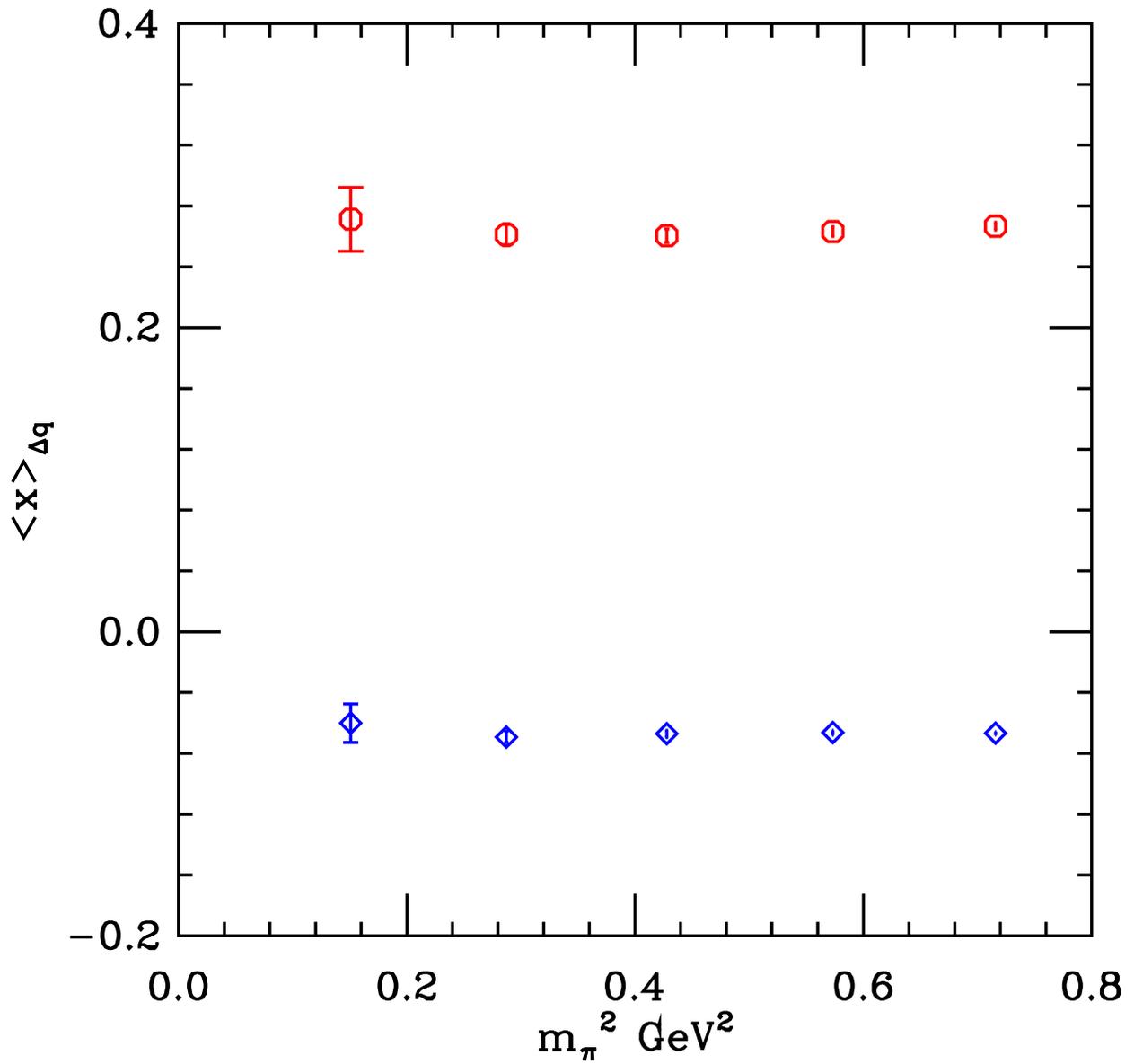}
\end{center}
\caption{The bare helicity distribution. Octagons are the up quark contributions and diamonds are the down quark contributions. Disconnected diagrams are not included.}
\label{fig:XDq}
\end{figure}

\begin{figure}
\begin{center}
\includegraphics[width=\textwidth]{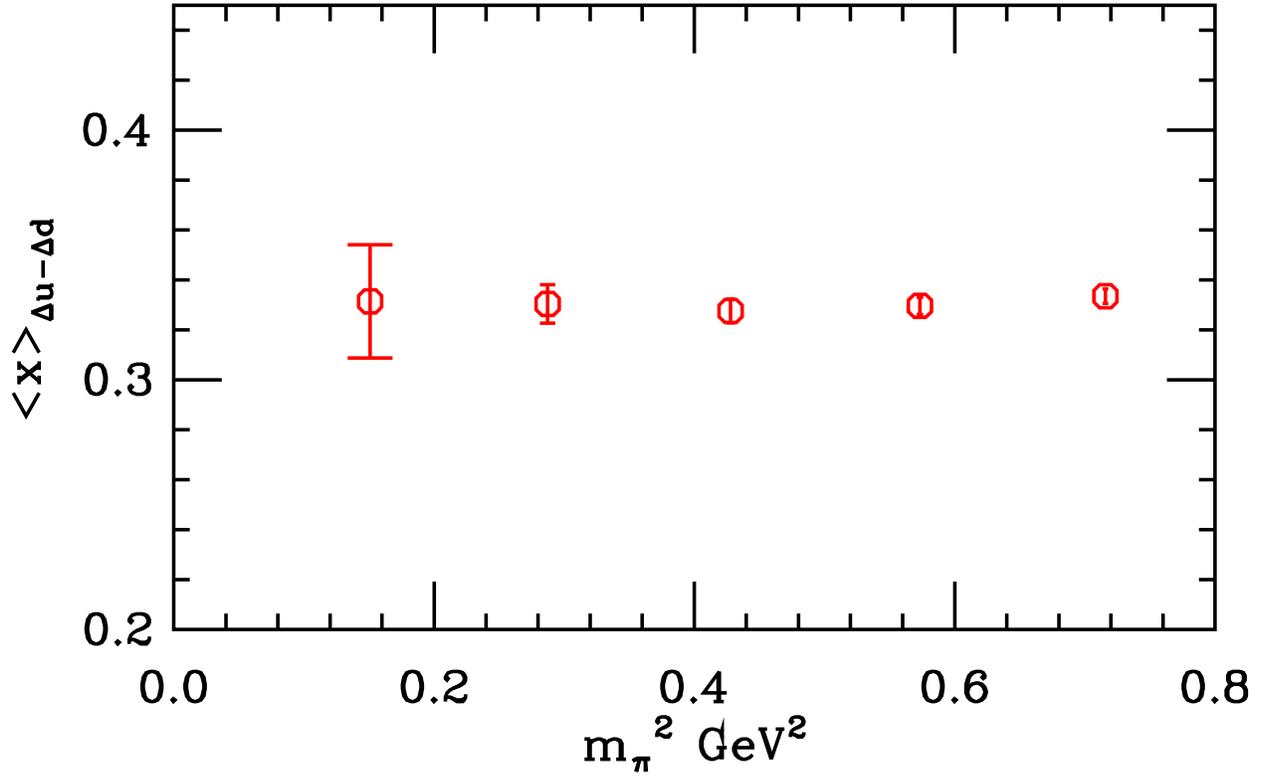}
\end{center}
\caption{The bare flavor non-singlet helicity distribution.}
\label{fig:XDq_ns}
\end{figure}

\begin{figure}
\begin{center}
\includegraphics[width=\textwidth]{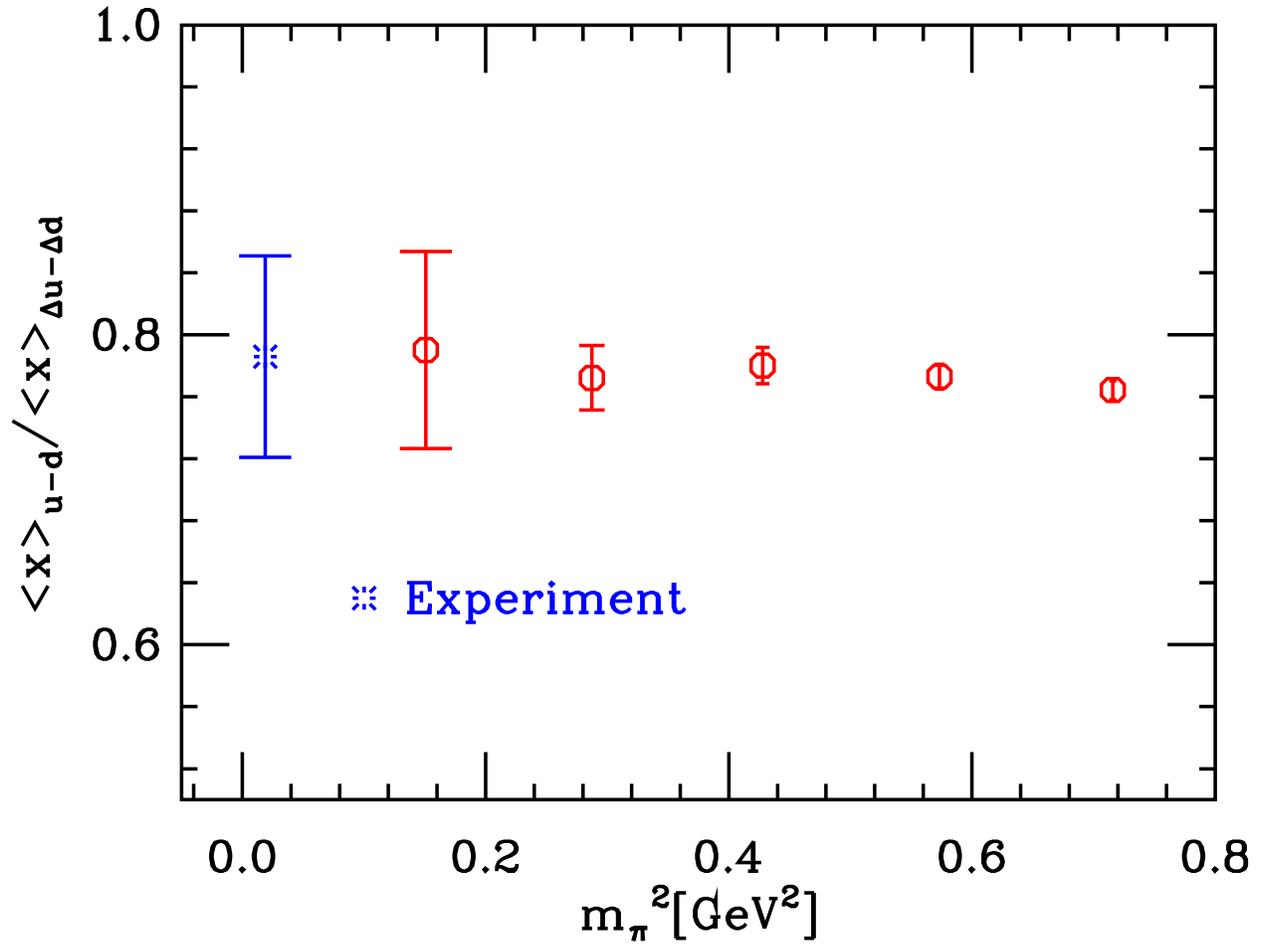}
\end{center}
\caption{The ratio of the  flavor non-singlet momentum fraction to the helicity distribution (octagons). The experimental expectation is marked by the burst symbol.}
\label{fig:Xq_o_XDq}
\end{figure}

\begin{figure}
\begin{center}
\includegraphics[width=\textwidth]{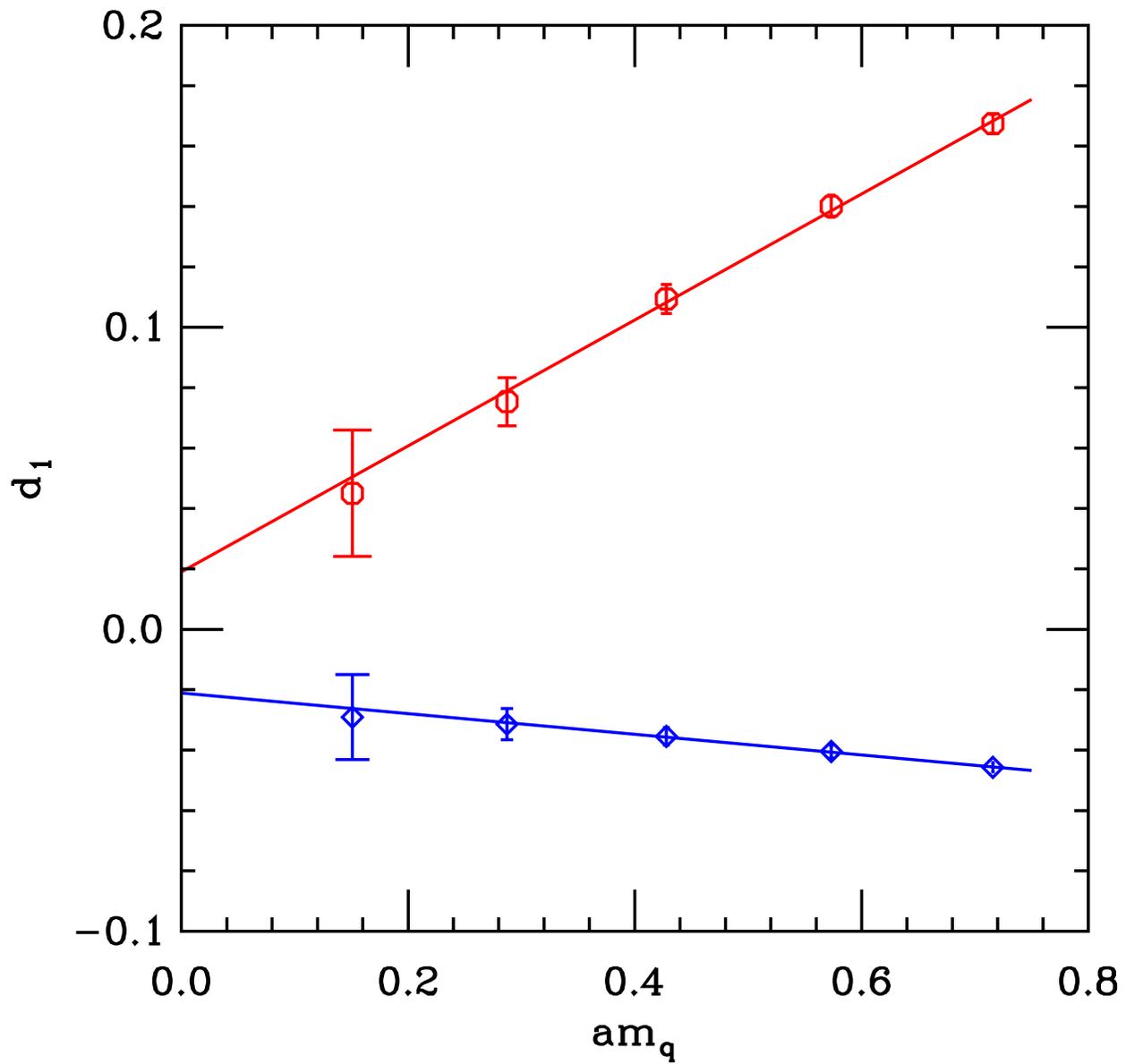}
\end{center}
\caption{The bare $d_1$. Octagons are the up quark contributions and diamonds are the down quark contributions. Disconnected diagrams are not included.}
\label{fig:d1}
\end{figure}

\begin{figure}
\begin{center}
\includegraphics[width=\textwidth]{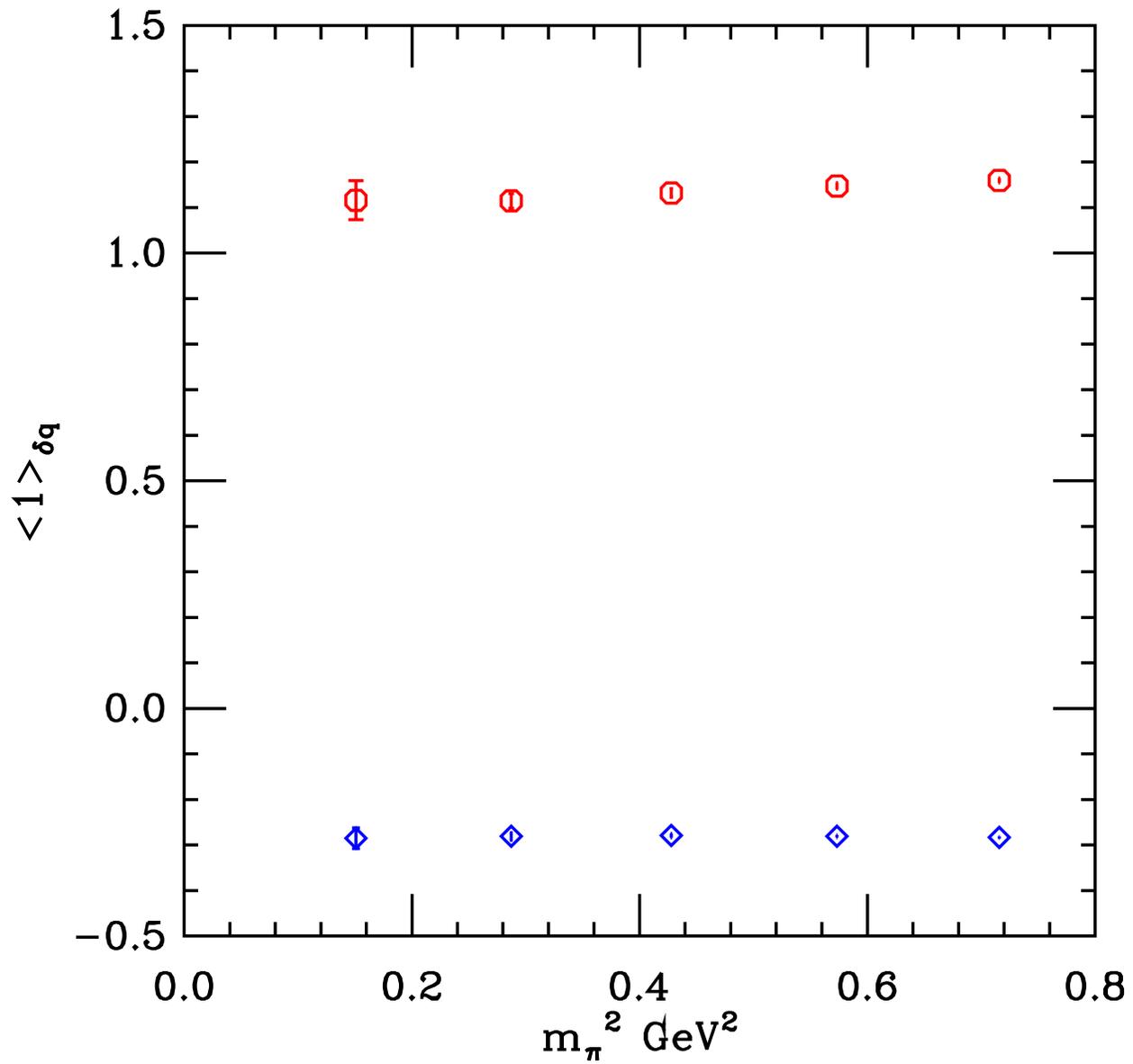}
\end{center}
\caption{The bare transversity. Octagons are the up quark contributions and diamonds are the down quark contributions. Disconnected diagrams are not included.}
\label{fig:1dq}
\end{figure}

\begin{figure}
\begin{center}
\includegraphics[width=\textwidth]{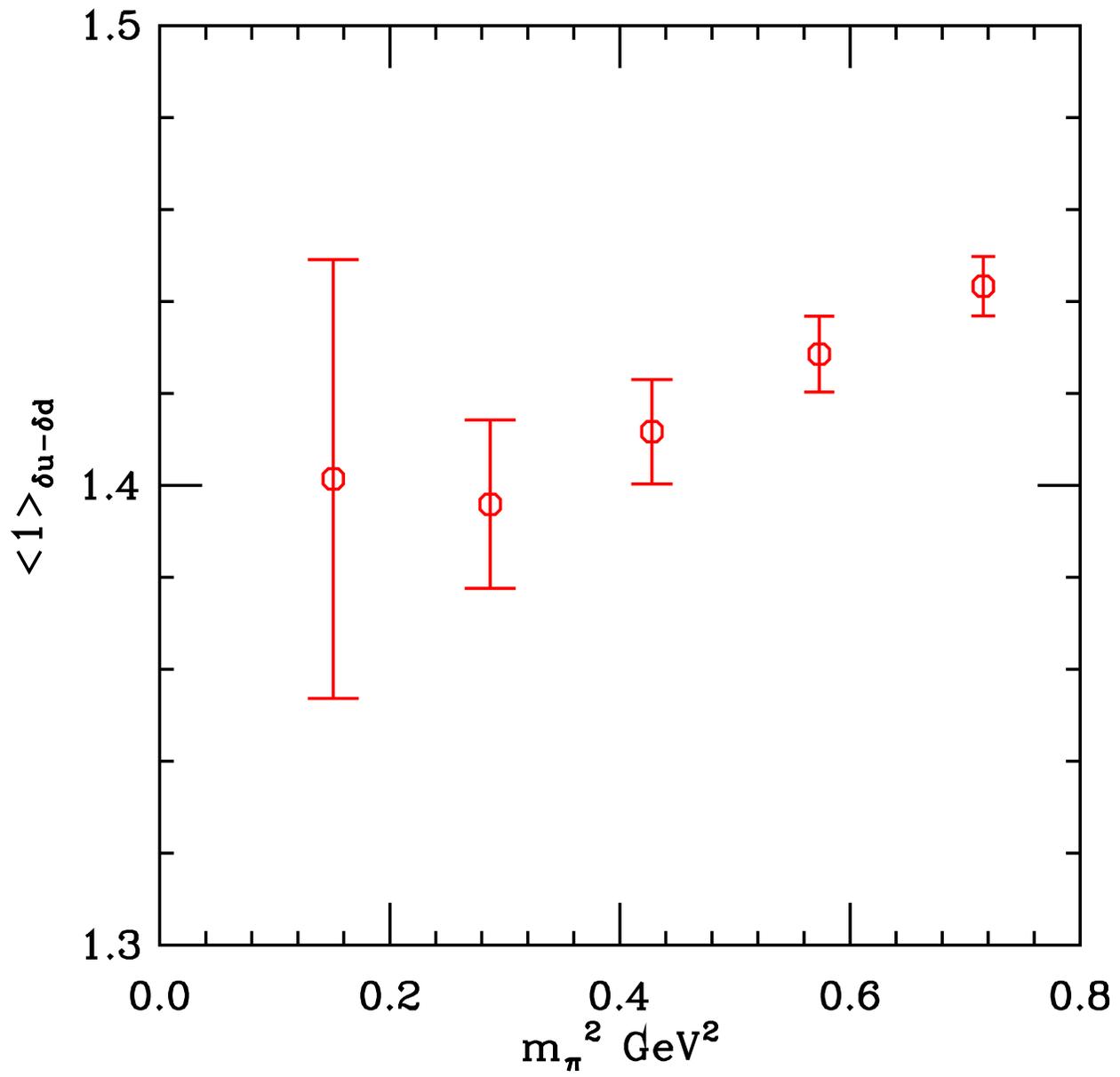}
\end{center}
\caption{The bare flavor non-singlet transversity.}
\label{fig:1dq_ns}
\end{figure}

\begin{figure}
\begin{center}
\includegraphics[width=\textwidth]{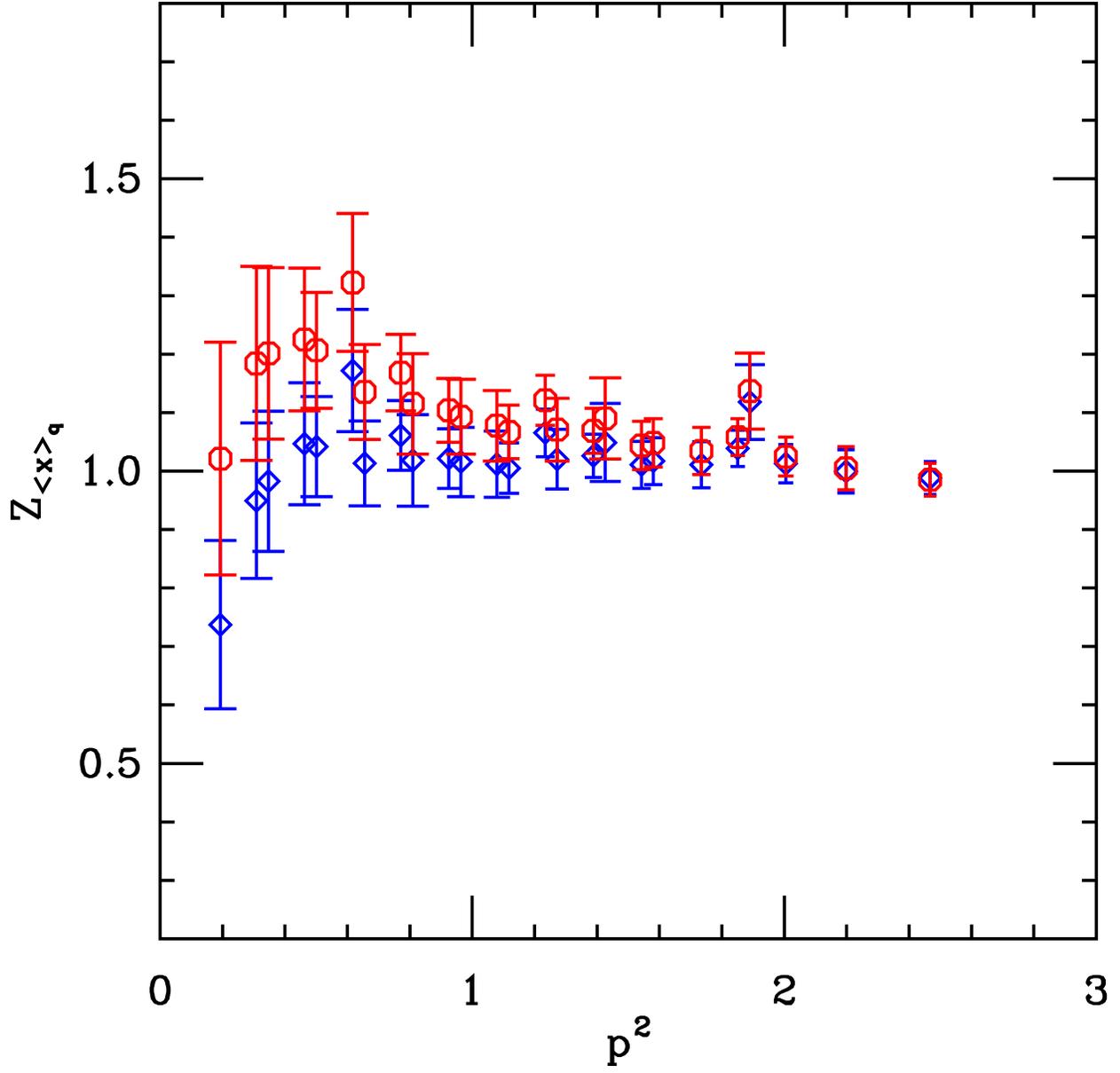}
\end{center}
\caption{Renormalization constant for the momentum fraction. The diamonds are the renormalization group invariant points. The octagons are the raw data.}
\label{fig:NPR-Xq}
\end{figure}

\begin{figure}
\begin{center}
\includegraphics[width=\textwidth]{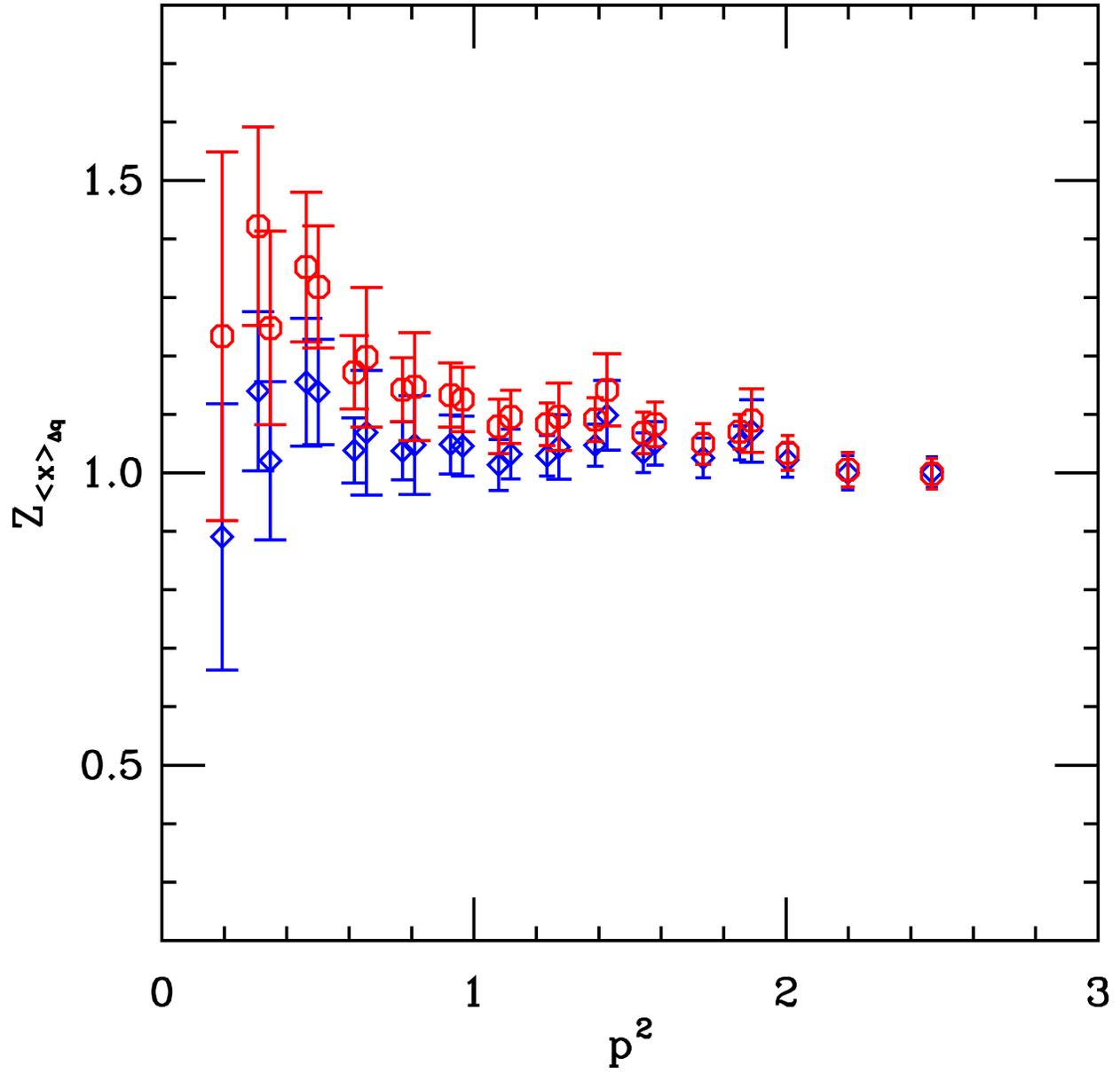}
\end{center}
\caption{Renormalization constant for the helicity distribution. The diamonds
are the renormalization group invariant points. The octagons are the raw data.}
\label{fig:NPR-XDq}
\end{figure}

\begin{figure}
\begin{center}
\includegraphics[width=\textwidth]{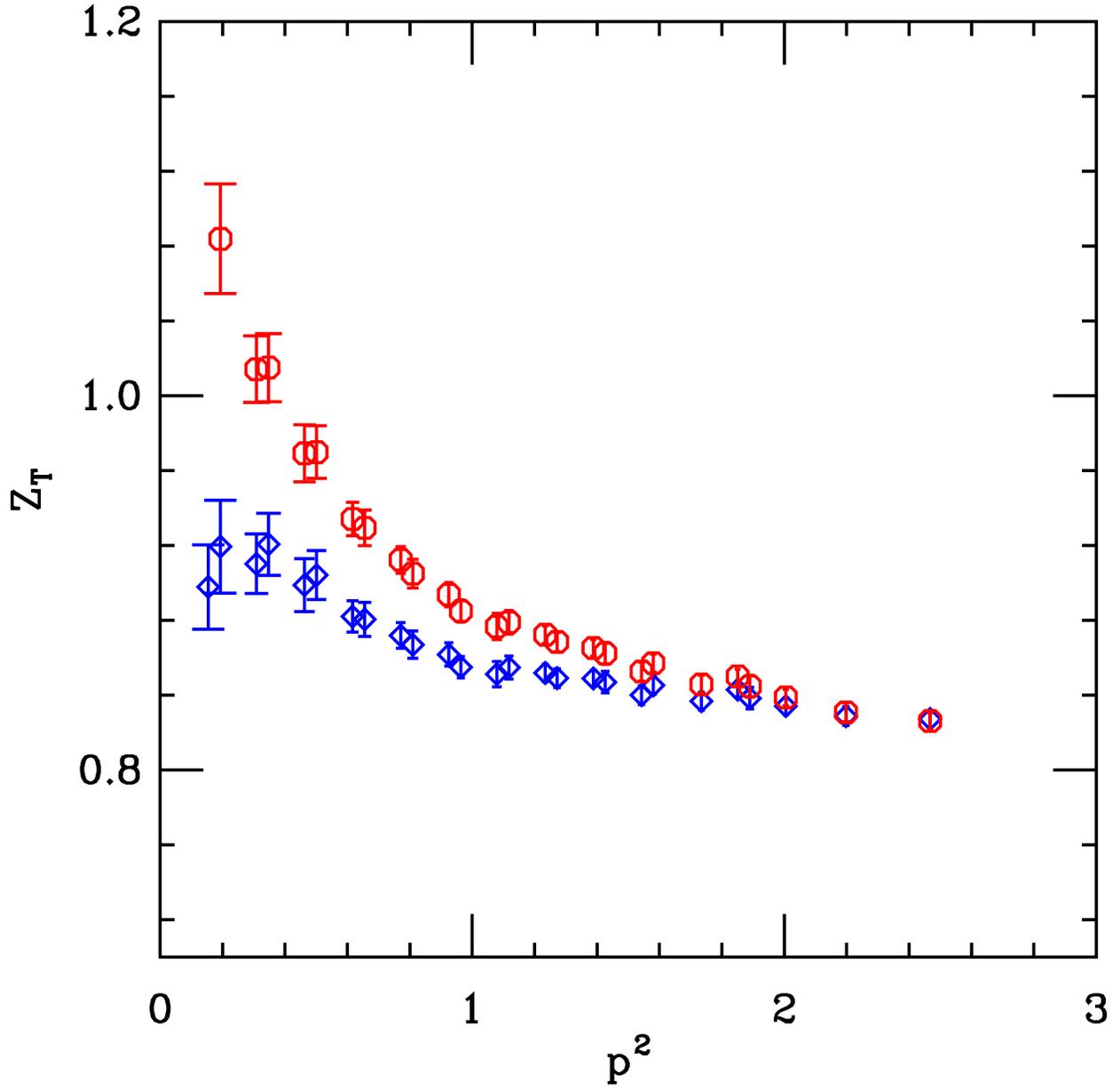}
\end{center}
\caption{Renormalization constant for the transversity. The diamonds
are the renormalization group invariant points. The octagons are the raw data.}
\label{fig:NPR-1dq}
\end{figure}

\begin{figure}
\begin{center}
\includegraphics[width=\textwidth]{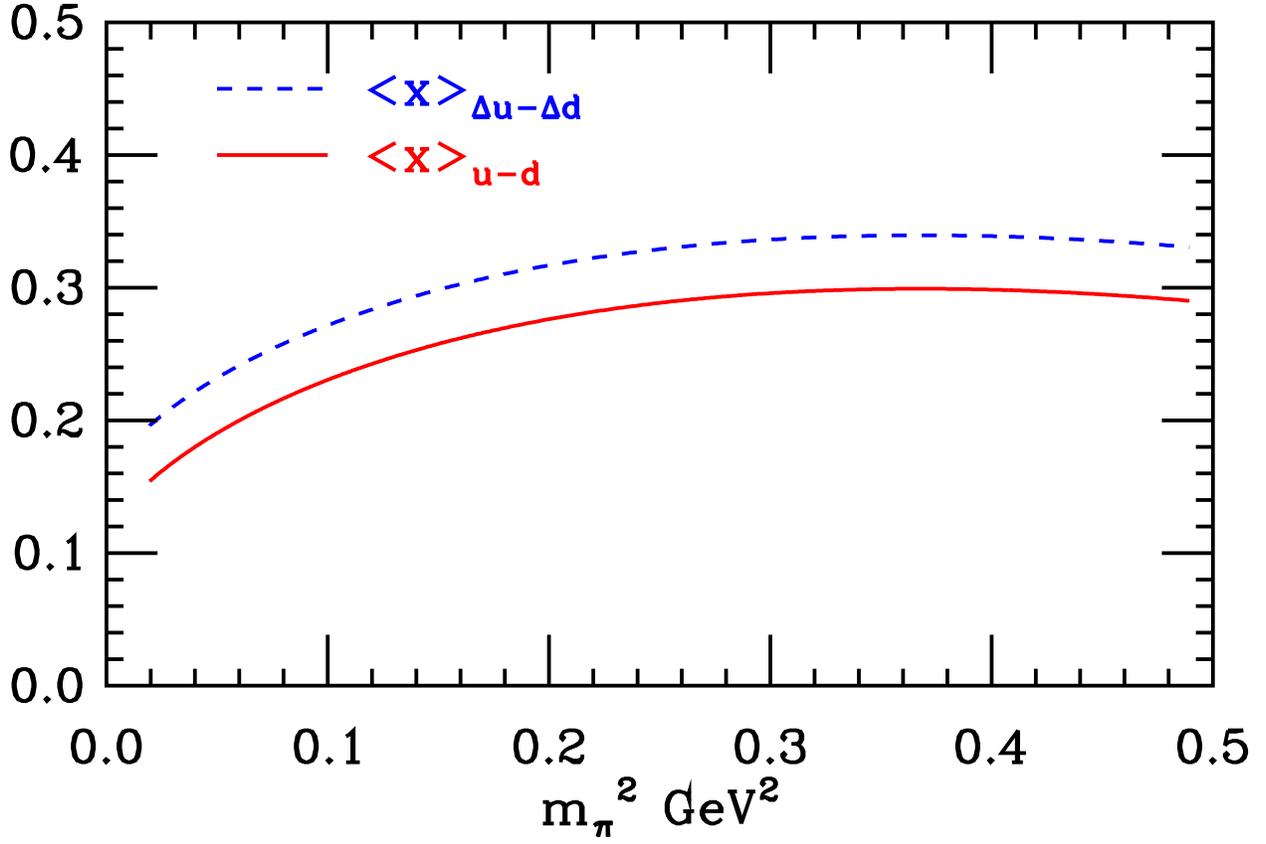}
\end{center}
\caption{The leading chiral logarithm dependence
for the first moment of the helicity and the momentum fraction. The curves are normalized so that at the physical point the experimental result is recovered.}
\label{fig:ChiPT}
\end{figure}

\begin{figure}
\begin{center}
\includegraphics[width=\textwidth]{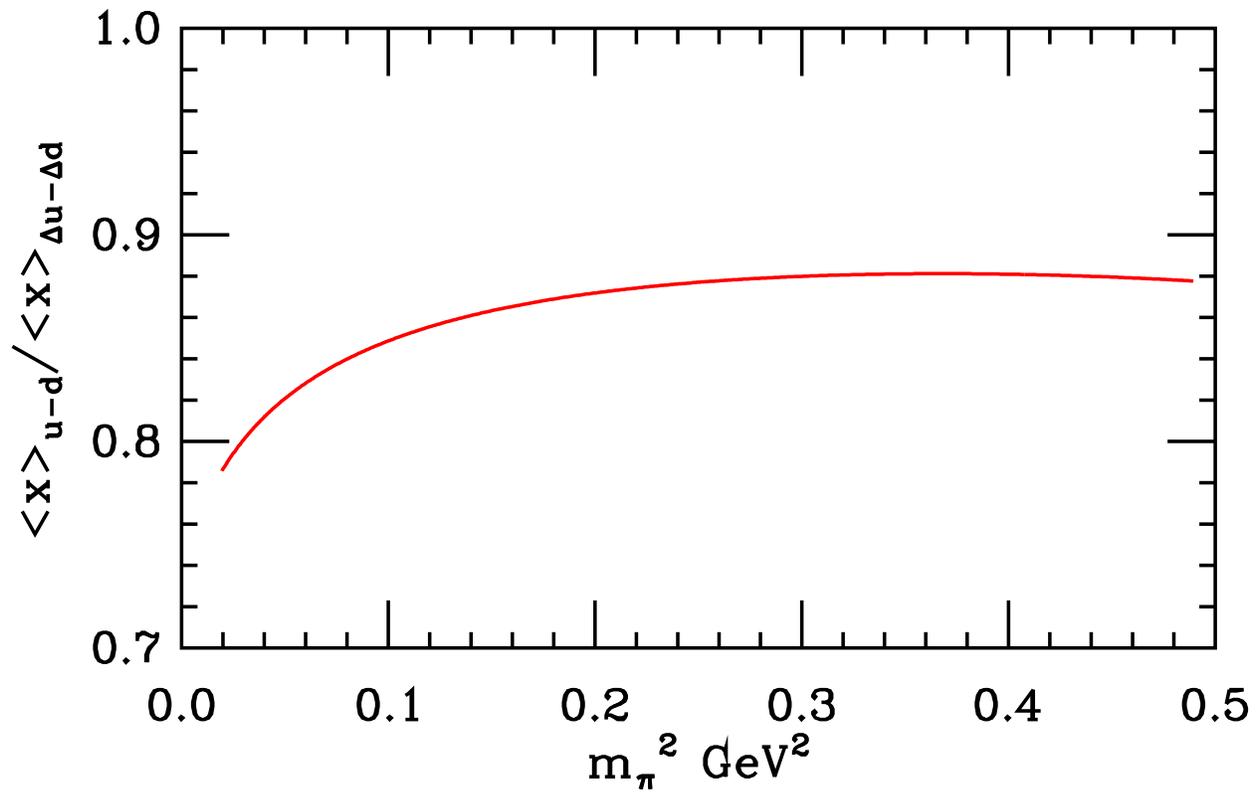}
\end{center}
\caption{The leading chiral logarithm dependence
for the ratio of the momentum fraction to the first moment of the helicity. The curve is normalized so that at the physical point the experimental result is recovered.}
\label{fig:ChiRatio}
\end{figure}

\clearpage
\newpage

\fi

\end{document}